\documentclass[fleqn,usenatbib]{mnras}

\usepackage{amsmath}
\usepackage{amssymb}
\usepackage{graphicx}
\usepackage{mathptmx}
\usepackage{mathrsfs}
\usepackage{txfonts}
\usepackage[T1]{fontenc}

\newcommand{\Aa}[2]{A_{\alpha,#1}^{#2}}
\newcommand{\ain}{a_\mathrm{in}}
\newcommand{\bahamasxl}{$\nu w$CDM}
\newcommand{\cambcode}{{\tt CAMB}}
\newcommand{\classcode}{{\tt CLASS}}
\newcommand{\cnu}{c_\nu}
\newcommand{\concept}{{\tt CO}{$\mathcal N$}{\tt CEPT}}
\newcommand{\cosmicenu}{{\tt Cosmic-E$\nu$}}
\newcommand{\DDcb}{\Delta^2_{\rm cb}}
\newcommand{\DDeh}{\Delta^2_{\rm EH}}
\newcommand{\DDehL}{\Delta^2_{{\rm EH},L}}
\newcommand{\DDehnu}{\Delta^2_{{\rm EH},\nu}}
\newcommand{\DDm}{\Delta^2_{\rm m}}
\newcommand{\DDnu}{\Delta^2_\nu}
\newcommand{\DDnuL}{\Delta^2_L}
\newcommand{\delD}{\delta^{({\rm D})}}
\newcommand{\delK}{\delta^{({\rm K})}}
\newcommand{\Dscr}{{\mathscr D}}
\newcommand{\EQ}[1]{Eq.~(\ref{#1})}
\newcommand{\EQS}[2]{Eqs.~(\ref{#1}-\ref{#2})}

\newcommand{\FFTM}{{\tt FlowsForTheMasses}}

\newcommand{\Hc}{{\mathcal H}}
\newcommand{\Hco}{{\mathcal H}_0}
\newcommand{\Ia}[2]{I_{\alpha, #1}^{#2}}

\newcommand{\kfs}{k_\mathrm{FS}}

\newcommand{\kthr}{k_{\rm st}}
\newcommand{\lamu}[1]{\lambda_{{\rm U},#1}}
\newcommand{\lamuh}[1]{{\hat\lambda}_{{\rm U},#1}}
\newcommand{\lamw}[1]{\lambda_{{\rm W},#1}}
\newcommand{\lamwh}[1]{{\hat\lambda}_{{\rm W},#1}}
\newcommand{\MFLR}{{\tt MuFLR}}
\newcommand{\Mnu}{M_\nu}
\newcommand{\mua}{\mu_\alpha}

\newcommand{\NC}{N_{\rm C}}
\newcommand{\Nkz}{N_{kz}}
\newcommand{\NM}{N_{\rm M}}
\newcommand{\Nmc}{N_{\rm MCMC}}
\newcommand{\Nmu}{N_\mu}
\newcommand{\Nmuai}{N_{\mu,AI}}
\newcommand{\Nmunl}{N_{\mu,\mathrm{NL}}}
\newcommand{\NPC}{N_{\rm PC}}
\newcommand{\Ntau}{N_\tau}

\newcommand{\Obo}{\Omega_{\mathrm{b},0}}
\newcommand{\Ocb}{\Omega_\mathrm{cb}}
\newcommand{\Ocbo}{\Omega_{\mathrm{cb},0}}

\newcommand{\Om}{\Omega_\mathrm{m}}

\newcommand{\Omo}{\Omega_{\mathrm{m},0}}
\newcommand{\On}{\Omega_\nu}

\newcommand{\Ono}{\Omega_{\nu,0}}
\newcommand{\Pa}[2]{P_{\alpha #1}^{#2}}
\newcommand{\PLeg}{{\mathcal P}}
\newcommand{\Pscr}{{\mathscr P}}

\newcommand{\Rnu}{R_\nu}
\newcommand{\Rscr}{{\mathscr R}}
\newcommand{\SEPIA}{{\tt SEPIA}}
\newcommand{\Tnuo}{T_{\nu,0}}
\newcommand{\va}{v_\alpha}
\newcommand{\Xia}[2]{\Xi_{\alpha #1}^{#2}}
\newcommand{\Xita}[2]{{\tilde \Xi}_{\alpha #1}^{#2}}

\title{Cosmic-E$\nu$: An emulator for the non-linear neutrino power spectrum}
\author[Amol Upadhye, {\em et al.}]{%
Amol Upadhye,$^{1,2,3}$ 
Juliana Kwan,$^{2,4}$
Ian G.~McCarthy,$^2$
Jaime Salcido,$^2$
Kelly R.~Moran,$^5$
\newauthor
Earl Lawrence,$^5$
Yvonne Y.~Y.~Wong$^3$
\\
$^1$ South-Western Institute for Astronomy Research, Yunnan University, Kunming 650500, People's Republic of China\\
$^2$ Astrophysics Research Institute, Liverpool John Moores University, 146 Brownlow Hill, Liverpool L3 5RF, United Kingdom\\
$^3$ Sydney Consortium for Particle Physics and Cosmology, School of Physics, The University of New South Wales,  Sydney NSW 2052, Australia\\
$^4$ Department of Applied Mathematics and Theoretical Physics, University of Cambridge, Cambridge CB3 0WA, United Kingdom\\
$^5$ Statistical Sciences Group, CCS Division, Los Alamos National Laboratory, Los Alamos, NM 87545
}

\begin{document}
\label{firstpage}
\pagerange{\pageref{firstpage}--\pageref{lastpage}}
\maketitle

\begin{abstract}
  Cosmology is poised to measure the neutrino mass sum $\Mnu$ and has identified several smaller-scale observables sensitive to neutrinos, necessitating accurate predictions of neutrino clustering over a wide range of length scales.  The \FFTM{} non-linear perturbation theory for the massive neutrino power spectrum, $\DDnu(k)$, agrees with its companion N-body simulation at the $10\%-15\%$ level for $k \leq 1~h/$Mpc.  Building upon the Mira-Titan IV emulator for the cold matter, we use \FFTM{} to construct an emulator for $\DDnu(k)$ covering a large range of cosmological parameters and neutrino fractions $\Ono h^2 \leq 0.01$, which corresponds to $\Mnu \leq 0.93$~eV.  Consistent with \FFTM{} at the $3.5\%$ level, it returns a power spectrum in milliseconds.  Ranking the neutrinos by initial momenta, we also emulate the power spectra of momentum deciles, providing information about their perturbed distribution function.  Comparing a $\Mnu=0.15$~eV model to a wide range of N-body simulation methods, we find agreement to $3\%$ for $k \leq 3\kfs = 0.17~h/$Mpc and to $19\%$ for $k \leq 0.4~h/$Mpc.   We find that the enhancement factor, the ratio of $\DDnu(k)$ to its linear-response equivalent, is most strongly correlated with $\Ono h^2$, and also with the clustering amplitude $\sigma_8$.  Furthermore, non-linearities enhance the free-streaming-limit scaling $\partial \log(\DDnu/\DDm) / \partial \log(\Mnu)$ beyond its linear value of $4$, increasing the $\Mnu$-sensitivity of the small-scale neutrino density.
\end{abstract}

%%%%%%%%%%%%%%%%%%%%%%%%%%%%%%%%%%%%%%%%%%%%%%%%%%%%%%%%%%%%%%%%%%%%%%%%%%%%%%%%
\section{Introduction}%%%%%%%%%%%%%%%%%%%%%%%%%%%%%%%%%%%%%%%%%%%%%%%%%%%%%%%%%%
\label{sec:int}%%%%%%%%%%%%%%%%%%%%%%%%%%%%%%%%%%%%%%%%%%%%%%%%%%%%%%%%%%%%%%%%%

Cosmology will measure the neutrino mass sum $\Mnu = \sum m_\nu$, one of the final unmeasured parameters of the Standard Model of particle physics, over the next several years, assuming that the dark energy is a cosmological constant~\citep{Audren:2012vy,Chudaykin:2019ock}.  Future space-based experiments will provide completely independent bounds on $\Mnu$~\citep{Petracca:2015alr,Lin:2022aro}.  However, both forecasts and analyses of current data are consistent with a weakening of the $\Mnu$ bound by a factor of $\approx 3$ when the dark energy equation of state is allowed to vary with time~\citep{Font-Ribera:2013rwa,Upadhye:2017hdl,DiValentino:2019dzu}.  Additionally, neutrinos and dark energy may play roles in the resolution of persistent tensions in measurements of the cosmic expansion and clustering amplitude~\citep{Leauthaud:2016jdb,Bohringer:2016fcq,Poulin:2018zxs,Gogoi:2020qif,DiValentino:2021imh,Mccarthy:2017yqf,McCarthy:2023ism}.

On the particle physics side, persistent anomalies in the neutrino sector motivate  models containing additional ``sterile'' neutrinos as well as non-standard neutrino interactions~\citep{Denton:2021czb,MiniBooNE:2022emn,Alvarez-Ruso:2021dna}.  Furthermore, other hot dark matter (HDM) species such as axions could mimic the cosmological effects of massive neutrinos~\citep{Giare:2021cqr,DEramo:2022nvb,DiValentino:2022edq}.  A quantitative understanding of neutrino clustering in the non-linear regime will prove invaluable for breaking these degeneracies.

In cosmology, several neutrino clustering signatures deep in the non-linear regime have been identified and quantified through N-body simulations.  These include "wakes" of neutrinos streaming coherently past halos~\citep{Zhu:2014qma,Inman:2015pfa}; an odd-parity contribution to the angular momentum field of galaxies~\citep{Yu:2018llx}; neutrino-dark-matter relative velocities~\citep{Zhu:2013tma,Inman:2016prk,Zhu:2019kzb,Zhou:2021sgl}; modifications to the halo mass function~\citep{Costanzi:2013bha,Yu:2016yfe,Biswas:2019uhy,Bocquet:2020tes,Ryu:2022npy};  a neutrino contribution to the scale-dependent bias of dark matter halos~\citep{LoVerde:2013lta,LoVerde:2014pxa,Chiang:2017vuk,Chiang:2018laa,Banerjee:2019omr}; and inhomogeneities in the cosmic neutrino background detectable in laboratory searches~\citep{PTOLEMY:2018jst,PTOLEMY:2019hkd,KATRIN:2022kkv}.  Since the systematic bias associated with these phenomena are substantially different from those of joint analyses of large-scale cosmic surveys, some of them may be decisive to a convincing detection of massive neutrinos.

Accurate theoretical modeling of non-linear neutrino clustering effects, possibly through an extension of the halo model to neutrino clustering, will require fast and reliable calculations of the neutrino power spectrum.  A recent code comparison by the Euclid collaboration, \citet{Euclid:2022qde}, tested a wide range of simulation methods (specifically, \citealt{SWIFT:2023dix,Teyssier:2001cp,Mauland:2023eax,Adamek:2015eda,Adamek:2016zes,Adamek:2017uiq,Beck:2015qva,Marin-Gilabert:2022ggx,Springel:2005mi,Springel:2008cc,Springel:2020plp,Dakin:2017idt,Dakin:2021ivb}) and found agreement at the $30\%-40\%$ level over the range $k \leq 1~h/$Mpc of wave numbers.  However, simulations are computationally expensive, motivating an exploration of alternative methods with comparable accuracies.

Linear perturbative calculations of massive neutrino clustering can be carried out to high precision by tracking the evolution of the neutrino distribution function~\citep{Ma:1995ey} in both position and momentum.  An alternative approach introduced by \citet{Dupuy:2013jaa,Dupuy:2014vea,Dupuy:2015ega} discretizes the Fermi-Dirac distribution of initial neutrino velocities.  Each neutrino fluid, defined by its initial velocity, can then be treated as a separate fluid individually obeying the continuity and Euler equations of fluid dynamics.  This approach is Lagrangian in momentum space, since neutrinos cannot move from one initial-velocity bin to another.  Since non-linear cosmological perturbation theory begins with the continuity and Euler equations, this momentum-Lagrangian method is a natural starting point for non-linear neutrino perturbation theories.

\FFTM{}, the first non-linear perturbative power spectrum calculation for free-streaming particles such as massive neutrinos, began with precisely this approach~\citep{Chen:2022cgw}.  Since a fluid with non-zero initial velocity $\vec v$ has a preferred direction $\hat v$, \citet{Chen:2022cgw} began by extending the Time-Renormalization Group perturbation theory of \citet{Pietroni:2008jx,Lesgourgues:2009am} to fluids with homogeneous initial velocities.  Their Fourier-space clustering depends not only upon the magnitude of the Fourier vector $\vec k$, but also its angle with $\hat v$, whose cosine is $\mu = \hat k \cdot \hat v$.

Expanding the density contrast and velocity divergence in Legendre polynomials in $\mu$, \citet{Chen:2022cgw} showed that the mode-coupling integrals of non-linear perturbation theory couple different Legendre moments, drastically increasing the computational cost.  However, by applying Fast Fourier Transform (FFT) techniques introduced by \citet{McEwen:2016fjn}, \citet{Schmittfull:2016jsw}, and \citet{Fang:2016wcf} to its mode-coupling integrals, \citet{Chen:2022cgw} was able to accelerate them by more than two orders of magnitude.  The resulting \FFTM{} perturbation theory can compute a non-linear massive neutrino power spectrum with reasonable accuracy settings on a standard desktop computer.\footnote{The \FFTM{} perturbation theory code is publicly available at {\tt github.com/upadhye/FlowsForTheMasses}~.}  %Here, we further develop \FFTM{} to improve its numerical stability at high $\Mnu$.

Though the computational cost of \FFTM{} is much lower than that of N-body neutrino simulations, it remains somewhat high; the fifty-flow production runs of~\citet{Chen:2022cgw} each took about a day on a $32$-core machine.  A machine learning technique known as emulation, introduced into cosmology by \citet{Heitmann:2008eq}, \citet{Heitmann:2009cu}, and \citet{Lawrence:2009uk}, is ideal for quickly approximating expensive functions.  Emulation begins with a training set of  evaluations of an expensive function (in our case the \FFTM{} neutrino power spectrum) the size of which is determined by computational budget and required level of accuracy. To mitigate the computational expense associated with emulating multivariate data, the spectra are represented via a principal component (PC) decomposition.  Gaussian process (GP) models are then used to model the cosmology-dependent PC weights, enabling fast prediction at new cosmologies.

Our goal in this work is an emulator of the \FFTM{} non-linear neutrino power spectrum.  Furthermore, since \FFTM{} already divides neutrinos by their initial momenta and tracks the evolution of each one separately, our training set includes individual neutrino momentum deciles at no extra computational cost.  For both the $z\leq 2$ CDM+baryon power spectra and the emulator design points, we use the Mira-Titan IV (MT4) emulator of \citet{Moran:2022iwe}.  We demonstrate that our emulator, \cosmicenu{}, precisely reproduces \FFTM{} to $<3.5\%$ for $10^{-3}~h/{\rm Mpc} \leq k \leq 1~h/$Mpc and $0 \leq z \leq 3$.~\footnote{\cosmicenu{} is publicly available at {\tt github.com/upadhye/Cosmic-Enu}~.}

Next, we compare \cosmicenu{} to a range of N-body neutrino simulations.  We begin with the Euclid code comparison of \citet{Euclid:2022qde} for $\Mnu=0.15$~eV, which runs simulations with a variety of resolutions, box sizes, and massive neutrino implementations.  Compared with their highest-resolution {\tt SWIFT} simulation of \citet{Schaller:2023hzn}, \cosmicenu{} is accurate to $3\%$ up to $k = 3 \kfs = 0.17~h/$Mpc, $19\%$ to $k=0.4~h/$Mpc, and $49\%$ to $k=1~h/$Mpc, all of which are somewhat larger than but comparable to the scatter among very different simulation methods.  This qualitative picture is unchanged when $\Mnu$ is raised to $0.3$~eV and $0.6$~eV in the \citet{Euclid:2022qde} simulations, and when the dark energy is allowed to vary rapidly in one of our own simulations (described below).

Finally, we employ \cosmicenu{} to study enhancements to the ratios $\DDnu / \DDnu[\rm LR]$ and $\DDnu / \DDm$ due to the non-linear clustering of massive neutrinos. The first of these, the non-linear enhancement of neutrino clustering relative to linear response (LR), was considered in  \citet{Chen:2022cgw}.  After confirming the accuracy of \cosmicenu{} for this quantity, we quantify the sensitivity of $\DDnu / \DDnu[\rm LR]$ to each of the eight cosmological parameters, showing that $\Ono h^2$ is by far the most significant for determining the neutrino clustering enhancement, followed by $\Obo h^2$, $\Omo h^2$, and $\sigma_8$.  The second ratio, $\DDnu / \DDm$, was shown by \citet{Ringwald:2004np} and \citet{Wong:2008ws} to scale as the fourth power of $\Mnu$, hence $\Ono h^2$, for LR neutrinos in the free-streaming limit.  We confirm this result in the linear case, then show that non-linear corrections enhance this scaling relation.  For example, at $k=1~h/$Mpc and $\Ono h^2 = 0.002$, $\DDnu / \DDm \sim \Mnu^{4.5}$.  

This study is organized as follows.  Section~\ref{sec:bkg} briefly describes the emulation procedure and the \FFTM{} perturbation theory.  Our emulator training set is assembled in Sec.~\ref{sec:tra} after improving the high-$\Mnu$ numerical stability of \FFTM{}.  Section~\ref{sec:emu} constructs the \cosmicenu{} emulator and quantifies its accuracy with respect to \FFTM{}.  \cosmicenu{} is then compared with a wide variety of N-body simulation methods in Sec.~\ref{sec:nbd}.  Finally, Sec.~\ref{sec:enh} quantifies the non-linear enhancements to the  $\DDnu / \DDnu[\rm LR]$ and $\DDnu / \DDm$ ratios, and Sec.~\ref{sec:con} concludes.

%%%%%%%%%%%%%%%%%%%%%%%%%%%%%%%%%%%%%%%%%%%%%%%%%%%%%%%%%%%%%%%%%%%%%%%%%%%%%%%%
\section{Background}%%%%%%%%%%%%%%%%%%%%%%%%%%%%%%%%%%%%%%%%%%%%%%%%%%%%%%%%%%%%
\label{sec:bkg}%%%%%%%%%%%%%%%%%%%%%%%%%%%%%%%%%%%%%%%%%%%%%%%%%%%%%%%%%%%%%%%%%

%-------------------------------------------------------------------------------
\subsection{Emulation}%---------------------------------------------------------
\label{subsec:bkg:emulation}

A thorough discussion of emulation in cosmology may be found in~\citet{Heitmann:2009cu}.  Here, we briefly summarize their procedure, with slight differences in notation.

Suppose that we wish to approximate a dimensionless function $\Pscr(k,z,\vec C)$ of the wave number $k$, redshift $z$, and cosmological parameters $\vec C$.  This may be proportional to the power spectrum itself, or a function of the power spectrum chosen to reduce its dynamic range.  At each of $m \in [0,\NM)$ cosmological models defined by parameters $\vec C^*_m$, we are given $\Pscr$ at each of $i \in [0,\Nkz)$ points $(k_i,z_i)$, that is, $\Pscr^*_{im} = \Pscr(k_i,z_i,\vec C^*_m)$.  We seek an approximation of the form
\begin{eqnarray}
  \Pscr(k_i,z_i,\vec C)
  &\approx&
  \mu^*(k_i,z_i) + \sigma^* \sum_{j=0}^{\NPC-1} w_j(\vec C\,) \, \phi_j(k_i,z_i)
  \label{e:Pscr_expansion}
  \\
  \mu^*(k_i,z_i)
  &:=&
  \frac{1}{\NM} \sum_{m=0}^{\NM-1} \Pscr^*_{im}
  \\
  (\sigma^*)^2
  &:=&
  \frac{1}{\NM \Nkz}
  \sum_{m=0}^{\NM-1} \sum_{i=0}^{\Nkz-1} \left(\Pscr^*_{im} - \mu^*(k_i,z_i)\right)^2
\end{eqnarray}
where $\mu_i^*$ is the mean input $\Pscr^*_{im}$ across cosmologies, the $\phi_j$ are a set of $\NPC$ orthogonal basis functions to be defined below, and $w_j(\vec C\,)$ are the corresponding basis weights.  
If the number of bases $\NPC$ is chosen equal to the number of training models $\NM$ minus one\footnote{Including the mean term reduces the remaining degrees of freedom in the model by one.}, then the approximation of \EQ{e:Pscr_expansion} can be made exact for the $\NM$ input models, though we will typically choose $\NPC$ smaller than this.

Let $\Dscr^*_{im} = (\Pscr^*_{im} - \mu^*_i)/\sigma^*$, so $\Dscr^*_{im}$ for fixed $i$ has zero mean by construction, and $\Dscr^*_{im}$ is an $\Nkz \times \NM$ matrix.  By means of a compact singular value decomposition, we may write
\begin{equation}
  \Dscr^*_{im}
  =
  \sum_{m'=0}^{\NM-1} \sum_{m''=0}^{\NM-1} U_{im'} D_{m'm''} (V^T)_{m''m}
\end{equation}
where $U$ is an $\Nkz \times \NM$ orthogonal matrix, $D$ an $\NM \times \NM$ diagonal matrix, and $V$ an $\NM\times\NM$ orthonormal matrix.  In terms of these three matrices, we may write the PC basis functions $\phi_j(k_i,z_i)$ and the weights of the input data $w^*_{jm}$ as
\begin{eqnarray}
  \phi_{ij}
  &:=&
  \phi_j(k_i,z_i) = \frac{1}{\sqrt{\NM}} \sum_{m=0}^{\NM-1} U_{im} D_{mj}
  \\
  w^*_{jm}
  &=&
  \sqrt{\NM} V_{mj}
\end{eqnarray}
where the $j$ index may be truncated to $j < \NPC$ for any chosen $\NPC < \NM$.  Since the functions $\phi_j(k,z)$ may be interpolated from the $\phi_{ij}$ using standard methods, our remaining task is to model the weight functions $w_j(\vec C\,)$ using $w^*_{jm}$.

We model each $w_j(\vec C\,)$ using a GP.  A GP is an infinite dimensional generalization of a multivariate Gaussian distribution, in which any finite set of random variables is defined to follow a multivariate Gaussian distribution specified by a mean function and a covariance function~\citep{Williams:2006gp}.  We define the GP over each $w_j(\vec C\,)$ to have a mean function of 0 and a Gaussian correlation function $R_j(\vec C, \vec C', \vec \beta_j)$. This correlation function is specified by a set of correlation hyperparameters $\beta_{j\ell}$, one for each of the $\NC$ cosmological parameters of $\vec C$:
\begin{equation}
  R_j(\vec C, \vec C', \vec \beta_j)
  =
  \prod_{\ell=0}^{\NC-1} \exp\left(-\beta_{j \ell} (C_\ell - C'_\ell)^2\right)
  \label{e:def_R}
\end{equation}

The input data weights $w^*_{jm}$ are now assumed to arise from the following hyperparameter-dependent probability distribution:
\begin{eqnarray}
  p(w^*_j|\lamw{j}, \lamu{j}, \vec \beta_j)
  &=&
  \frac{
    \exp\left[-\frac{\lamw{j}}{2}
      \sum_{m=0}^{\NM-1} \sum_{n=0}^{\NM-1}
      w^*_{jm} \Rscr(\vec\beta)^{-1}_{jmn} w^*_{jn}\right]
    }{\sqrt{\det{\Rscr}} (2\pi/\lamw{j})^{\NM/2}}\quad
  \\
  \Rscr_{jmn}(\lamw{j},\lamu{j},\vec\beta_j)
  &=&
  \lamu{j}^{-1} R_j(\vec C^*_m, \vec C^*_n, \vec\beta_j) + \lamw{j}^{-1} \delK_{mn},
  \label{e:Rscr}
\end{eqnarray}
where $\delK_{mn}$ is the Kronecker delta.  \citet{Heitmann:2009cu} has included an additional set of hyperparameters, a scaling term $\lamu{j}$ and a ``nugget'' term $\lamw{j}$. The former scales the correlation function into a covariance function, while the latter accommodates slight numerical fluctuations in the computation of $\Pscr^*_{im}$.   For each principal component $j$, we thus have one $\lamu{j}$, one $\lamw{j}$, and $\NC$ different $\beta_{j\ell}$, for a total of $\NC+2$ hyperparameters.

Since we are given $w^*_{jm}$ but not the hyperparameter values, our next step is to find the hyperparameter values most consistent with $w^*_{jm}$.  We do so using a $50,000$-step Markov chain Monte Carlo sampling through the \SEPIA{} code of \citet{SEPIA}\footnote{\SEPIA{} is publicly available at {\tt github.com/lanl/SEPIA }~.}, using the default hyperparameter priors and bounds defined in \SEPIA{}.  Let $\lamuh{j}$, $\lamwh{j}$, and $\hat{\vec{\beta}}_j$ be the posterior mean values of these hyperparameters. 

We have now arrived at our goal, a predictive model for the weights $w_j(\vec C\,)$ in \EQ{e:Pscr_expansion} for a given cosmology $\vec C$. Let $\hat{\Theta}=\{\hat{\vec{\beta}}_j, \lamuh{j}, \lamwh{j}\}$. Using these optimal hyperparameter values, we specify the conditional Gaussian probability distribution of each of the weights as follows:
\begin{eqnarray}
    w_j(\vec C\,|\,w_j^{\,*}, \hat{\Theta}\,) &\sim& N(\bar W_j(\vec C\,), \bar \Sigma_j(\vec C\,)), \label{e:hat_p} \\
    \bar W_j(\vec C\,|\,w_j^{\,*}, \hat{\Theta}\,) &=& [\vec r_j^{\:*}(\vec C\,)]^T [\hat\Rscr_j^*]^{-1} \vec w_j^{\,*}, \label{e:bar_Wj} \\
    \bar \Sigma_j(\vec C\,|\,\hat{\Theta}\,) &=& r_j(\vec C\,) - [\vec r_j^{\:*}(\vec C\,)]^T [\hat\Rscr_j^*]^{-1} \vec r_j^{\:*}(\vec C\,), \label{e:bar_Sigj} \\
    r_{jm}^{\:*}(\vec C\,|\,\hat{\Theta}\,) &=& \lamuh{j}^{-1} R_j(\vec C^*_m, \vec C, \hat{\vec{\beta}}_j). \label{e:def_rjm}
\end{eqnarray}
Here, $\vec w_j^{\,*}$ is the $N_M \times 1$ vector of observed weights, $\hat\Rscr_j^*$ is the $N_M \times N_M$ matrix of realizations of \EQ{e:Rscr} evaluated at the optimal values of the hyperparameters, $\vec r_j^{\:*}(\vec C\,)$ is an $N_M \times 1$ vector having $m$th entry $r_{jm}^{\:*}(\vec C\,)$, and $r_j(\vec C\,) = \lamuh{j}^{-1} R_j(\vec C, \vec C, \hat{\vec{\beta}}_j) + \lamwh{j}$. Equation~(\ref{e:hat_p}) indicates that $w_j$ is drawn from a normalized Gaussian distribution of mean $\bar W_j(\vec C\,)$ and standard deviation $\bar \Sigma_j(\vec C\,)$.  As in \cite{Heitmann:2009cu}, we use $\bar W_j(\vec C\,)$ as our emulated prediction of $w_j(\vec C\,)$. We see from \EQ{e:bar_Wj} that this mean value is a weighted average of the observed weights $\vec w_j^{\,*}$, with weights determined by the covariance between the observed and predictive cosmological parameters.

Thus far we have assumed the initial data $\Pscr^*_{im}$ to be given.  Choosing the $\NM$ input models used to train the emulator is a separate problem known as emulator design.  The goal of emulator design is to cover the given parameter space to a specified accuracy using the smallest number $\NM$ of input cosmological models, since each model is computationally expensive.  A simple grid in parameter space is one of the least efficient designs for this purpose. \citet{Heitmann:2009cu,Heitmann:2015xma,Lawrence:2017ost,Moran:2022iwe} employ efficient space-filling Latin hypercube designs or similar nested, space-filling lattices; the latter is useful when runs are to be done in ``batches'' so that interim analyses may be performed as partial results are available. The training sets we construct, described in Sec. \ref{subsec:tra:emulator_design}, are built on the design choices made by these authors. 

The batched MT4 emulator design allows for the addition of more design points to improve its accuracy.  We leave this possibility open for future work.  However, we will see in Sec.~\ref{subsec:emu:tests_of_cosmicenu} that even with the existing MT4 design, \cosmicenu{} attains an accuracy of $\approx 3.5\%$, which is subdominant to the $14\%$ error in \FFTM{} itself, as measured by \citet{Chen:2022cgw}.  Thus we do not pursue here the possibility of including additional design points.

%-------------------------------------------------------------------------------
\subsection{Non-linear perturbation theory for neutrinos}%----------------------
\label{subsec:bkg:nonlinear_perturbation_theory_for_neutrinos}

We are particularly interested in the non-linear growth of the neutrino density perturbations, which occurs at late times and at scales well within the Hubble horizon.  Thus, to excellent approximation, we assume that all matter in the Universe obeys the scalarized non-relativistic continuity, Euler, and Poisson equations in a box expanding uniformly at a rate given by the time-dependent Hubble parameter.  General Relativistic clustering including vector and tensor perturbations, as well as multiple fluids, has been considered previously (e.g., \citealt{Hwang:2005wb,Hwang:2005xt,Hwang:2007ni,Hwang:2012aa,Hwang:2012ra,Hwang:2015prq,Jeong:2010ag,Gong:2017tev,Yoo:2014sfa,Yoo:2014kpa,Magi:2022nfy,Adamek:2014xba,Adamek:2016zes,Adamek:2017uiq,Fidler:2015npa,Fidler:2016tir,Fidler:2017pnb,Fidler:2018bkg}).  Generalizing these results to multiple fluids with different non-zero initial velocities is beyond the scope of the present study, as well as unnecessary to our goal of accuracy in the non-linear regime.  Furthermore, following \citep{Moran:2022iwe}, we restrict our consideration to spatially-flat cosmologies.

The chief difficulty in applying the continuity and Euler fluid equations to neutrinos is that, in the Eulerian fluid description, neutrinos are not fluids.  At each spatial point, neutrinos have a velocity dispersion arising from their initial Fermi-Dirac distribution; the number of degrees of freedom required to describe neutrino perturbations is therefore technically infinite.  \citet{Fuhrer:2014zka} devised a Eulerian fluid-like perturbation theory for the neutrino bispectrum by absorbing the infinite degrees of freedom into temporally non-local couplings, while \citet{Garny:2020ilv,Garny:2022fsh} combined a linear treatment of free streaming with non-linear corrections to the density and velocity monopole perturbations. 

On the other hand, the approach of \citet{Dupuy:2013jaa,Dupuy:2014vea,Dupuy:2015ega} and \citet{Chen:2020bdf,Chen:2022cgw},  which we describe in below, is to formulate a neutrino perturbation theory that is Lagrangian in momentum space.  Let $\vec \tau$ be the lower-index three-momentum in the limit of an unperturbed universe, $P_i^{(0)}$, which is time-independent.  The Fermi-Dirac distribution may be binned by the magnitude $\tau = |\vec \tau|$ of this momentum, allowing us to approximate the neutrino population using $\Ntau$ momenta, $\tau_\alpha$, for $\alpha \in [0,\Ntau)$.

For a given momentum vector $\vec\tau_\alpha$ in bin $\alpha$, the set of neutrinos with initial momentum $\vec\tau_\alpha$ has no thermal velocity dispersion.  Thus it behaves as a fluid obeying the continuity and Euler equations.  Spatial isotropy implies that these equations depend upon the direction of $\vec\tau_\alpha$ only through its angle with respect to the Fourier vector $\vec k$, that is, through $\mua := \vec k \cdot \vec\tau_\alpha / (k\tau_\alpha)$.  \citet{Chen:2020bdf} demonstrates that this $\mua$-dependence can be expanded in $\Nmu$ Legendre polynomials $\PLeg_\ell(\mua)$, using an appropriate boundary term at $\ell=\Nmu-1$, with an error of order $\Nmu^{-2}$.  Furthermore, for given $\tau_\alpha$, all neutrino fluids with initial momenta $\vec\tau$ such that $|\vec\tau|=\tau_\alpha$ obey the same fluid equations.  We use the term ``flow'' for this entire set of fluids.

The subhorizon non-relativistic linear theory of \citet{Chen:2020bdf} began with dimensionless scalar perturbations to the density, $\delta_\alpha(\vec x) = (\rho_\alpha(x)-\bar\rho_\alpha)/\bar\rho_\alpha$, and the momentum divergence, $\theta_\alpha(\vec x) = -\vec\nabla \cdot \vec P / (m_\nu a \Hc)$, with $m_\nu$ the neutrino mass and $\Hc$ the conformal Hubble expansion rate.  Since linear theory allows us to choose an arbitrary normalization for the perturbation variables, we normalize them to the square roots of their corresponding power spectra,
\begin{eqnarray}
  \delta_\alpha^{\vec k}
  &=&
  \sum_\ell (-i)^\ell  \PLeg_\ell(\mua) \delta_{\alpha\ell}^k
  \quad\textrm{and}\quad
  \theta_\alpha^{\vec k} = \sum_\ell  (-i)^\ell \PLeg_\ell(\mua) \theta_{\alpha\ell}^k
  \label{e:delta_PLeg_exp}
  \\
  \Pa{00}{\vec k}
  &=&
  \sum_\ell \PLeg_\ell(\mua)^2 \delta_{\alpha\ell}^k  \delta_{\alpha\ell}^k
  =: \sum_\ell \PLeg_\ell(\mua)^2 \Pa{00\ell}{k}
  \label{e:P00_def}
  \\
  \Pa{11}{\vec k}
  &=&
  \sum_\ell \PLeg_\ell(\mua)^2 \theta_{\alpha\ell}^k  \theta_{\alpha\ell}^k
  =: \sum_\ell \PLeg_\ell(\mua)^2 \Pa{11\ell}{k}
  \label{e:P11_def}
\end{eqnarray}
where $\PLeg_\ell(x)$ is the Legendre polynomial of order $\ell$.  Above we have employed two conventions we will use henceforth.
\begin{enumerate}
\item Power spectrum indices $b$ and $c$ in $\Pa{bc}{}$ take the value $0$ for $\delta$ and $1$ for $\theta$, so, for example, $\Pa{00}{} = \Pa{\delta\delta}{}$.
\item Wave number superscripts denote a functional dependence, so that $\Pa{bc}{\vec k} = \Pa{bc}{}(\vec k)$ and $\delta_{\alpha\ell}^k = \delta_{\alpha\ell}(k)$.
\end{enumerate}

The perfect correlation between the random variables corresponding to $\delta_\alpha(\vec k)$ and $\theta_\alpha(\vec k)$ in linear theory breaks down beyond the linear order, leading the \FFTM{} perturbation theory of \citet{Chen:2022cgw} to introduce the quantities
\begin{equation}
  \chi_{\alpha\ell}^k
  :=
  1 - \frac{\Pa{01\ell}{k}}{\sqrt{\Pa{00\ell}{k}\Pa{11\ell}{k}}}
  \Rightarrow
  \Pa{01\ell}{k} = (1-\chi_{\alpha\ell}^k)\delta_{\alpha\ell}^k \theta_{\alpha\ell}^k.
  \label{e:chi_P01_def}
\end{equation}
Since $\Pa{01}{} = \Pa{10}{}$, \EQS{e:delta_PLeg_exp}{e:chi_P01_def} completely specify the power spectra $\Pa{bc}{\vec k}$ in terms of the perturbation variables.

In terms of the bispectrum integrals $\Ia{acd,bef,\ell}{k}$ of \citet{Chen:2022cgw}, the evolution of $\delta_{\alpha\ell}$, $\theta_{\alpha\ell}$, and $\chi_{\alpha\ell}$ is given by
\begin{eqnarray}
  (\delta_{\alpha\ell}^k)'
  &=&
  \tfrac{k\va}{\Hc}\left(\tfrac{\ell}{2\ell-1}\delta_{\alpha,\ell-1}^k
  - \tfrac{\ell+1}{2\ell+3}\delta_{\alpha,\ell+1}^k\right)
  + \theta_{\alpha\ell}^k
  + \tfrac{2}{\delta_{\alpha\ell}^k} \Ia{001,001,\ell}{k}\quad
  \label{e:eom_delta}
  \\
  (\theta_{\alpha\ell}^k)'
  &=&
  - \left(1+\tfrac{\Hc'}{\Hc}\right)\theta_{\alpha\ell}^k
  + \tfrac{k\va}{\Hc}\left(\tfrac{\ell}{2\ell-1}\theta_{\alpha,\ell-1}^k
  - \tfrac{\ell+1}{2\ell+3}\theta_{\alpha,\ell+1}^k\right)
  \ldots\nonumber\\
  &~&
  -\delK_{\ell0}\tfrac{k^2\Phi^k}{\Hc^2}
  + \tfrac{1}{\theta_{\alpha\ell}^k} \Ia{111,111,\ell}{k}\qquad
  \label{e:eom_theta}
  \\
  (\chi_{\alpha\ell}^k)'
  &=&
  \tfrac{2(1-\chi_{\alpha\ell}^k)}{(\delta_{\alpha\ell}^k)^2}\Ia{001,001,\ell}{k}
  + \tfrac{1-\chi_{\alpha\ell}^k}{(\theta_{\alpha\ell}^k)^2} \Ia{111,111,\ell}{k}
  \ldots\nonumber\\
  &~&
  - \tfrac{2}{\delta_{\alpha\ell}^k \theta_{\alpha\ell}^k} \Ia{001,101,\ell}{k}
  - \tfrac{1}{\delta_{\alpha\ell}^k \theta_{\alpha\ell}^k} \Ia{111,011,\ell}{k}.
  \label{e:eom_chi}
\end{eqnarray}
where $\va := \tau_\alpha / (m_\nu a)$ is the flow velocity, and primes denote derivatives with respect to $\eta := \log(a/\ain)$ for given initial scale factor $\ain$.  The gravitational potential $\Phi$ is given by the Poisson equation
\begin{equation}
  k^2\Phi^k
  =
  -\frac{3}{2}\Hc^2
  \left( \Ocb(\eta) \delta_\mathrm{cb}^k
  + \sum_{\alpha=0}^{N_\tau-1} \Omega_\alpha(\eta) \delta_{\alpha 0}^k \right)
  \label{e:eom_Phi}
\end{equation}
where $\delta_\mathrm{cb}$ is the density contrast of the CDM and baryons, treated as a single fluid, labelled cb; $\Ocb(\eta) = \Ocbo \Hco^2 / (a \Hc^2)$ is the time-dependent density fraction of this cb fluid; $\Ocbo$ its density fraction today; $\Omega_\alpha(\eta) = \Omega_{\alpha,0} \Hco^2 / (a \Hc^2(1-\va^2)^{1/2})$ the time-dependent density fraction of neutrino flow $\alpha$; and $\Omega_{\alpha,0}$ its density fraction today.

The bispectrum integrals $\Ia{acd,bef,\ell}{k}$ are defined and thoroughly studied in \citet{Chen:2022cgw}.  For our purposes, we may define them by their evolution equations
\begin{eqnarray}
    (\Ia{acd,bef,\ell}{k})'
  &=&
  -\Xia{bg\ell}{k} \Ia{acd,gef,\ell}{k}
  - \Xita{eg\ell}{k} \Ia{acd,bgf,\ell}{k}
  \ldots\nonumber\\
  &~&
  - \Xita{fg\ell}{k} \Ia{acd,beg,\ell}{k}
  + 2\Aa{acd,bef,\ell}{k}
  \label{e:eom_I}
  \\
  \Xia{ab\ell}{k}
  &=&
  \left[
    \begin{array}{cc}
      0 & -1 \\
      \frac{k^2 \Phi^k}{\Hc^2 \delta_{\alpha 0}^k}\delK_{\ell 0}~~
      & 1 \!+\! \frac{\Hc'}{\Hc}
    \end{array}
    \right]
  \ldots\nonumber\\
  &~&
  - \delK_{a0}\delK_{b0}
  \frac{k v_\alpha}{\Hc}
  \!\left(\!\frac{\ell}{2\ell-1}
  \frac{\delta_{\alpha,\ell-1}^{k}}{\delta_{\alpha\ell}^{k}}
  \!-\!\frac{\ell+1}{2\ell+3}
  \frac{\delta_{\alpha,\ell+1}^{k}}{\delta_{\alpha\ell}^{k}}
  \!\right)
  \ldots\nonumber\\
  &~&
  - \delK_{a1}\delK_{b1}
  \frac{k v_\alpha}{\Hc}
  \!\left(\!\frac{\ell}{2\ell-1}
  \frac{\theta_{\alpha,\ell-1}^{k}}{\theta_{\alpha\ell}^{k}}
  \!-\!\frac{\ell+1}{2\ell+3}
  \frac{\theta_{\alpha,\ell+1}^{k}}{\theta_{\alpha\ell}^{k}}
  \!\right)
  \label{e:def_Xia}
  \\
  \Xita{ab\ell}{k}
  &=&
  \left[
    \begin{array}{cc}
      0~~~ & -1 \\
      0~~~ & 1 + \frac{\Hc'}{\Hc}
    \end{array}
    \right]
  \label{e:def_Xita}
\end{eqnarray}
with initial conditions $\Ia{acd,bef,\ell}{k} = 2\Aa{acd,bef,\ell}{k}$ at $\eta=0$.  Here, the mode-coupling integrals are given by
\begin{eqnarray}
  \Aa{acd,bef}{\vec k}
  &:=&
  \!\!\int \!\!\!\tfrac{d^3q}{(2\pi)^3} \tfrac{d^3p}{(2\pi)^3}
  (2\pi)^3 \delD(\vec k - \vec p - \vec q)
  \gamma_{acd}^{\vec k \vec q \vec p}
  \Bigg[ \gamma_{bgh}^{\vec k \vec q \vec p} \Pa{ge}{\vec q} \Pa{hf}{\vec p}
  \ldots\nonumber\\
  &~&
  \quad
  + \gamma_{egh}^{\vec q,-\vec p,\vec k} \Pa{gf}{\vec p} \Pa{hb}{\vec k}
  + \gamma_{fgh}^{\vec p,\vec k,-\vec q} \Pa{gb}{\vec k} \Pa{he}{\vec q} \Bigg].
  \label{e:def_A}
  \\
  &=:&
  \sum_\ell  \PLeg_\ell(\mua)^2 \Aa{acd,bef,\ell}{k}
  \label{e:A_ell}
  \\
  \gamma_{001}^{\vec k \vec q \vec p}
  &=&
  \frac{(\vec q + \vec p)\cdot \vec p}{2p^2}
  ,\quad
  \gamma_{010}^{\vec k \vec q \vec p} = \gamma_{001}^{\vec k \vec p \vec q}
  \label{e:gamma}
  ,\quad
  \gamma_{111}^{\vec k \vec q \vec p}
  =
  \frac{(\vec q + \vec p)^2 \vec q \cdot \vec p}{2 q^2 p^2}
\end{eqnarray}
with all other $\gamma_{abc}$ vanishing.  In Eqs.~(\ref{e:eom_I},\ref{e:def_A}), we have assumed implicit summation over the indices $g$ and $h$ for compactness.  Numerical computation of these mode-coupling integrals $\Aa{acd,bef,\ell}{k}$ is the main computational expense of \FFTM{} perturbation theory.  \citet{Chen:2022cgw} shows that they may be reduced to Fast Fourier Transforms (FFTs) and then computed using the methods of \citet{Hamilton:1999uv,McEwen:2016fjn,Fang:2016wcf,Schmittfull:2016jsw,Upadhye:2017hdl}.

%%%%%%%%%%%%%%%%%%%%%%%%%%%%%%%%%%%%%%%%%%%%%%%%%%%%%%%%%%%%%%%%%%%%%%%%%%%%%%%%
\section{Training data set}%%%%%%%%%%%%%%%%%%%%%%%%%%%%%%%%%%%%%%%%%%%%%%%%%%%%%
\label{sec:tra}%%%%%%%%%%%%%%%%%%%%%%%%%%%%%%%%%%%%%%%%%%%%%%%%%%%%%%%%%%%%%%%%%

%-------------------------------------------------------------------------------
\subsection{Emulator design}%---------------------------------------------------
\label{subsec:tra:emulator_design}

We begin by constructing the training set $\Pscr^*_{im}$ upon which the emulator is built.  Emulator design is described broadly in \citet{Heitmann:2009cu}, and the particular design of the Mira-Titan IV (MT4) emulator upon which \cosmicenu{} is built is described in \citet{Heitmann:2015xma}, \citet{Lawrence:2017ost}, and \citet{Moran:2022iwe}.  The $111$ design points in cosmological parameter space, $\vec C^*_m$, are chosen to strike a balance between broad parameter coverage and a high density of points, necessary for achieving high accuracy.  Out of these, $\NM=101$ design points have non-zero neutrino masses, with physical density fractions $\Ono h^2$ ranging from $0.00017$ to $0.01$, corresponding to $\Mnu$ from $0.0158$~eV to $0.931$~eV.

Our strategy is to build upon the MT4 emulator design.  Since massless neutrinos are already described well by linear theory, \cosmicenu{} uses only these $\NM=101$ massive-neutrino points.  We build our emulator upon MT4 for two reasons.  Firstly, the MT4 design has already been optimized and thoroughly tested for a balance between breadth and accuracy.  Secondly, the MT4 CDM+baryon power spectrum is used as an input to the \FFTM{} neutrino perturbation theory.  Since MT4 is most accurate at its own design points, choosing this same design for \cosmicenu{} avoids the compounding of errors that would result from using the outputs of one emulator as the inputs for another.  Another potential error is the backreaction of enhanced neutrino clustering on the CDM+baryon power.  \citet{Chen:2020bdf} quantified the linear response backreaction to be $0.05\%$ for $\Ono h^2=0.01$, the largest value considered here, while Section~\ref{subsec:enh:neutrino_contribution_to_matter_power} argues that non-linear clustering only increases this by a factor of $\sim 3$.  Thus this backreaction is a negligible source of error for \cosmicenu{}.

\begin{table}
  \caption{
    Allowed ranges of each cosmological parameter in \cosmicenu{}.  As with 
    the MT4 emulator, we have assumed three degenerate-mass neutrinos.
    \label{t:MT4_range}
  }
  \begin{center}
    \begin{tabular}{c|cc}
      parameter          & minimum   & maximum \\
      \hline
      $\Omo h^2$         & $0.12$    & $0.155$ \\
      $\Obo h^2$         & $0.0215$  & $0.0235$ \\
      $\Ono h^2$         & $0.00017$ & $0.01$ \\
      $\sigma_8$         & $0.7$     & $0.9$ \\
      $h$                & $0.55$    & $0.85$ \\
      $n_{\rm s}$         & $0.85$    & $1.05$ \\
      $w_0$              & $-1.3$    & $-0.7$ \\
      $(-w_0-w_a)^{1/4}$  & $0.3$     & $1.29$
    \end{tabular}
  \end{center}
\end{table}

Table~\ref{t:MT4_range} lists the parameter ranges covered by the \cosmicenu{} emulator.  The allowed range of $\Ono h^2$ values is determined by the set of $101$ massive-neutrino MT4 models.  Its lower bound of $0.00017$ is over three times smaller than the lower bound imposed by laboratory oscillation experiments~\citep{deSalas:2017kay,Capozzi:2018ubv,Esteban:2020cvm}. In Sec.~\ref{subsec:emu:tests_of_cosmicenu} we will quantify \cosmicenu{} errors near this low-$\Ono h^2$ boundary. For still smaller values, the non-relativistic-neutrino approximation made by \FFTM{} becomes increasingly inaccurate, and we instead recommend the use of relativistic linear perturbation theories such as \classcode{}~\citep{Lesgourgues:2011re,Blas:2011rf,Lesgourgues:2011rh} and \cambcode{}~\citep{Lewis:1999bs,Lewis:2002ah}.  The upper bound on $\Ono h^2$, consistent with $\Mnu=0.931$~eV, allows for a broad exploration of the parameter space, which may be useful for finding solutions to the Hubble and $\sigma_8$ tensions~\citep{Mccarthy:2017yqf,DiValentino:2021imh,McCarthy:2023ism}.   As with the MT4 emulator, we have assumed three degenerate-mass neutrinos.

Allowed ranges on the remaining seven parameters are taken directly from the MT4 emulator.  We parameterize the dark energy equation of state as $w(z) = w_0 + w_a z / (1+z)$, following \citet{Chevallier:2000qy,Linder:2002et}. The final range in Table~\ref{t:MT4_range} implies that if $w_0=-1$, then $-1.77 \leq w_a \leq 0.99$.   Allowing significant variation in the dark energy equation of state is important for cosmological constraints, as \citet{Upadhye:2017hdl} showed that this variation weakens the neutrino mass bound by a factor of $\approx 3$.

%-------------------------------------------------------------------------------
\subsection{Stabilizing perturbation theory}%-----------------------------------
\label{subsec:tra:stabilizing_perturbation_theory}

Our next challenge is the high-$k$ numerical instability of the \FFTM{} perturbation theory.  The mode-coupling integral $\Aa{acd,bef,\ell}{k}$ at low $k$ and large $\ell$ rises sharply with $k$, increasing its dynamic range.  Since FFTs spread errors across the entire range, small numerical errors near the peak of $\Aa{acd,bef,\ell}{}$ lead to large fractional errors where $\Aa{acd,bef,\ell}{}$ is small.  These lead to instabilities, preventing integration of the equations of motion at high $k$.  Additionally, the computational cost of the full set of mode-coupling integrals $\Aa{acd,bef,\ell}{k}$ of \EQS{e:def_A}{e:A_ell} was shown by \citet{Chen:2022cgw} to scale as $\Nmu^6$, further motivating a truncation in the range of $\ell$.

In practice, \FFTM{} integrates $128$ values of the wave number, logarithmically distributed between $10^{-4}~h/$Mpc and $10~h/$Mpc, and sets a stability threshold $\kthr$ to $10~h/$Mpc at the beginning of integration.  The integration step size in $\eta$ is chosen dynamically.  Each time that high-$k$ numerical instabilities drive this step size below $10^{-6}$, the integrator discards the highest $k$ value, effectively lowering $\kthr$ by $9.1\%$, and then resumes integration for $k \leq \kthr$.  Since non-linear physics causes power to flow from low to high $k$, we may safely discard wave numbers above this stability threshold without affecting those below it.  We will see below that no significant noise or discontinuities affect the power spectrum in the range $k \leq \kthr$.  As an added precaution, we will require stability up to a threshold $\kthr$ that is $20\%$ larger than our largest wave number of interest, $1~h/$Mpc.

In order to stabilize \FFTM{} for its fiducial model, with $\Ono h^2 = 0.005$, \citet{Chen:2022cgw} truncated the power spectra $\Pa{bc\ell}{k}$ used to compute $\Aa{acd,bef}{\vec k}$ to $\ell < \Nmunl$, while allowing $\Nmunl \leq \ell < \Nmu$ elsewhere.  They demonstrated that $\Nmunl$ of $6$, $7$, and $8$, respectively agreed with their N-body neutrino simulations to $14\%$, $12\%$, and $10\%$ for $k \leq 1~h/$Mpc.  However, $\Nmunl$ of $9$ suffered from severe numerical instabilities preventing the equations of motion from being integrated to the present time.  Henceforth we fix $\Nmunl=6$, since its substantial reduction in computational expense relative to $7$ and $8$ only modestly decreases its accuracy.

Although the $\ell < \Nmunl$ truncation of \citet{Chen:2022cgw} sufficed to stabilize their $\Ono h^2 = 0.005$ model over the range $k \leq 3~h/$Mpc, we find that increasing the neutrino density fraction tends to exacerbate the numerical instabilities in \FFTM{}.  The MT4 emulator of \citet{Moran:2022iwe} allows $\Ono h^2$ to be twice as high as the fiducial model of \citet{Chen:2022cgw}.  Requiring stability up to $k=1.2~h/$Mpc, so as to allow for a buffer around our desired range $k \leq 1~h/$Mpc, we find that $19$ of the $101$ design models are numerically unstable (that is, have $\kthr < 1.2~h/$Mpc).  The mean and minimum $\Ono h^2$ values for these models are $0.0083$ and $0.0063$, respectively, so this instability is a high-$\Ono h^2$ problem.

As $\Ono h^2$, hence the neutrino mass sum, increases, flow velocities $\va$ decrease, gradually decoupling the neutrino density and velocity monopoles from higher Legendre moments $\ell$.  Thus we employ a more aggressive high-$\ell$ truncation in order to stabilize these high-$\Ono h^2$ models.  In addition to truncating the power spectrum used to compute mode-coupling integrals, we truncate the $\ell$ expansions of the mode-coupling integrals $\Aa{acd,bef,\ell}{k}$ and bispectrum integrals $\Ia{acd,bef,\ell}{k}$ themselves, $\ell < \Nmuai$.  The power truncation $\ell < \Nmunl$ itself implies an $\Nmuai$ of $2\Nmunl - 1$, or $11$ for $\Nmunl=6$, so we allow $\Nmuai$ to be reduced below this number.

\begin{figure}
  \includegraphics[width=85mm]{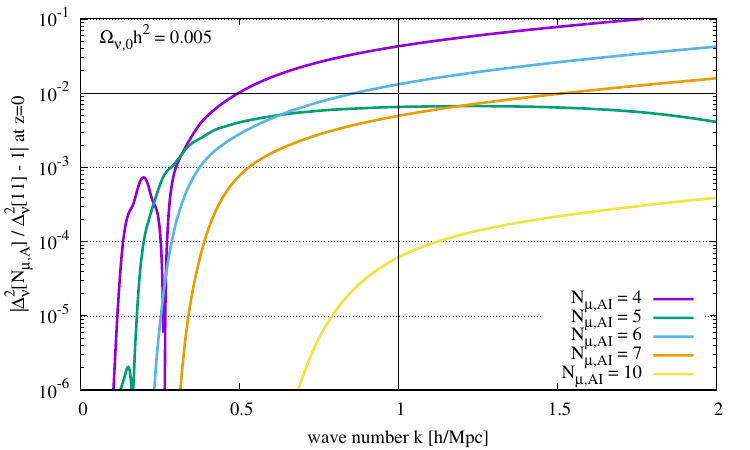}%
  
  \includegraphics[width=85mm]{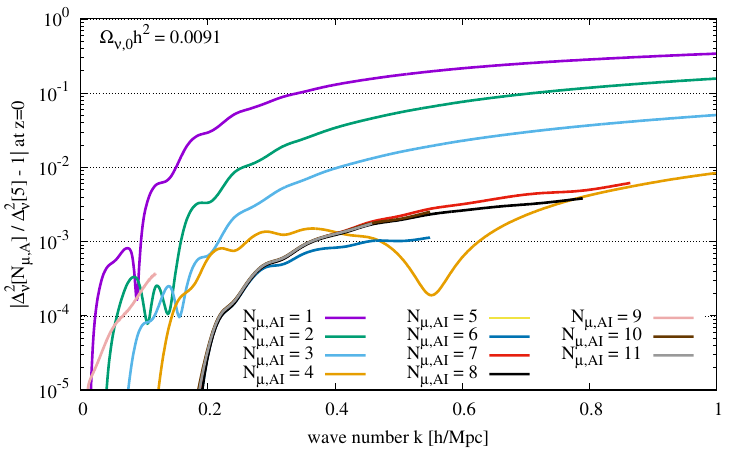}%
  \caption{
    Accuracy of truncation $\ell < \Nmuai$ in mode-coupling and bispectrum
    integrals.  (Top) $\Ono h^2 = 0.005$.  All $\Nmuai$ are stable up to
    $k = 2~h/$Mpc, so the power spectrum for each has been divided by that
    for $\Nmuai=11$, the maximum value.  (Bottom) MT4 design model with
    $\Ono h^2 = 0.0091$.  Each power spectrum has been divided by that for
    $\Nmuai=5$, the largest value for which the calculation is stable up to
    $k=1.2~h/$Mpc.
    \label{f:vary_Nmuai}
  }
\end{figure}

As an upper bound on the error associated with this truncation, consider the $\Ono h^2 = 0.005$ fiducial model of \citet{Chen:2022cgw}.  Figure~\ref{f:vary_Nmuai}~(Top) compares neutrino power spectra for several $\Nmuai$ values to the maximum value $2\Nmunl-1 = 11$.  In the range $k \leq 1~h/$Mpc, $\Nmuai=4$ is accurate to $4\%$, while the higher $\Nmuai$ considered are accurate to $1.3\%$ or better.

Figure~\ref{f:vary_Nmuai}~(Bottom) makes a similar comparison for one of the MT4 design points with $\Ono h^2 = 0.0091$.  Since power spectra with $\Nmuai>5$ cannot be stably integrated to $z=0$ all the way to $k=1.2~h/$Mpc, we use the $\Nmuai=5$ power spectrum for comparison.  Encouragingly, the $\Nmuai=4$ power spectrum agrees with this to better than $1\%$ at all $k \leq 1~h/$Mpc.  Also, even those higher-$\Nmuai$ power spectra with $\kthr < 1~h/$Mpc agree with the $\Nmuai=5$ power to better than $1\%$ across their entire stable ranges $k \leq \kthr$.  Evidently from the figure, our stabilization procedure discards higher $k$ early enough to prevent them from contaminating $k \leq \kthr$ at even the percent level.

Out of the nineteen MT4 design points requiring a mode-coupling truncation $\Nmuai < 2\Nmunl-1$, fifteen reach $k=1.2~h/$Mpc with $\Nmuai=5$, and the remaining four reach that wave number with $\Nmuai=4$.  These four $\Nmuai=4$ models all have $\Ono h^2 > 0.008$, with a mean $\Ono h^2$ value of $0.0092$.  All have stability thresholds $\kthr > 0.7~h/$Mpc when run with $\Nmuai=5$.  By comparison with Fig.~\ref{f:vary_Nmuai}~(Bottom), we may estimate their truncation error as $\sim 1\%$.  Another estimate of the truncation error, for all nineteen stabilized models, is the difference between the $\Nmuai=4$ and $\Nmuai=5$ power spectra, either to $k=1~h/$Mpc or to $\kthr$ if this is less than one.  By this measure, we find that one of these nineteen has a $2.5\%$ truncation error, while all others have $\leq 2\%$ and nine of them have $\leq 1\%$.  Since these nineteen represent nearly half of the forty design models with $\Ono h^2 \geq 0.0063$, we estimate that truncation leads to a $\approx 1\%$ power underestimate in that range.

%-------------------------------------------------------------------------------
\subsection{Reduced power spectrum}%--------------------------------------------
\label{subsec:tra:reduced_power_spectrum}

The power spectrum of massive neutrinos declines sharply below the free-streaming scale, $k \gg \kfs$, giving $\DDnu(k)$ a large dynamic range, which in turn makes it difficult to emulate.  Our strategy is to divide the neutrino power spectra, the total power as well as the single-decile power spectra, by quickly-calculable linear approximations.  Further reduction in the dynamic range is achieved by taking the natural logarithm of this power spectrum ratio, for each of the MT4 models, resulting in the training set $\Pscr^*_{im}$.

As a starting point for a fast linear approximation to the neutrino power spectra, we take the matter power spectra of Eisenstein and Hu \citep{Eisenstein:1997ik,Hu:1997vi,Eisenstein:1997jh}.  In the clustering limit, $k \ll \kfs$, neutrinos and cold matter cluster very similarly.  In the free-streaming limit, $k \gg \kfs$, the ratio of the neutrino and total matter density contrasts scales as $\kfs^2/k^2$, as shown by \citet{Ringwald:2004np} and \citet{Wong:2008ws}.  Those studies developed and tested an interpolation function, $\delta_\nu / \delta_{\rm m} \approx (1+k/\kfs)^{-2}$, that applies to linearly-clustering neutrinos, even when the CDM and baryons cluster non-linearly.

\citet{Chen:2020bdf,Chen:2020kxi} generalized the free-streaming scale of \citet{Ringwald:2004np} to individual neutrino flows through the replacement of their neutrino sound speed $\cnu$ by the flow velocity $v$,
\begin{equation}
  \kfs(a,v)^2 = \frac{3 \Om(a) \Hc(a)^2}{2 v(a)^2}
\end{equation}
where, in the non-relativistic approximation, $v(a) = \tau / (a m_\nu)$, implying that $\kfs(a,v) \propto \sqrt{a}$.  Thus our Eisenstein-Hu-like approximations to the total and decile neutrino power spectra are respectively
\begin{eqnarray}
  \DDehnu(k,a)
  &=&
  \DDeh(k,a) \left( 1 + \tfrac{k}{\kfs(a,\cnu)} \right)^{-4}
  ~\textrm{with}~
  \cnu^2 = \tfrac{3 \zeta(3) \Tnuo^2}{2 \log(2) m_\nu^2 a^2}\quad
  \\
  \DDehL(k,a)
  &=&
  \DDeh(k,a) \left( 1 + \tfrac{k}{\kfs(a,v_L)} \right)^{-4}
  ~\textrm{with}~
  v_L = \left< v_\alpha \right>_L
\end{eqnarray}
where $\left< v_\alpha \right>_L$ in the final line denotes an average over all flows $\alpha$ making up decile $L$, and $\zeta(x)$ is the Riemann zeta function.

\begin{figure}
  \includegraphics[width=85mm]{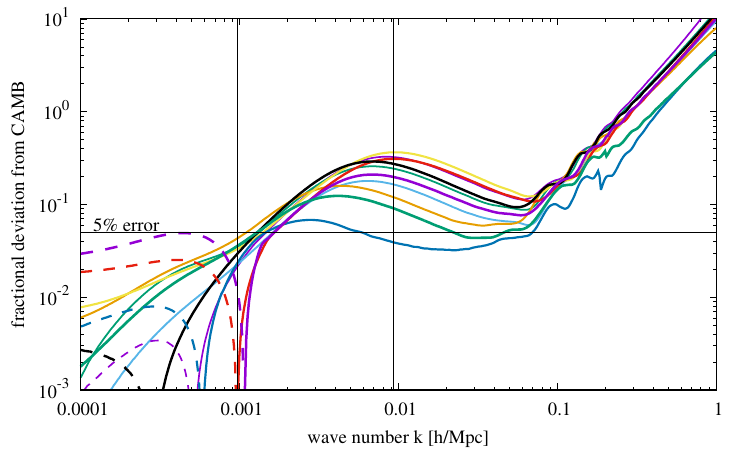}%
  \caption{
    Fractional difference between \FFTM{} and \cambcode{} for ten cosmologies spanning the parameter space, listed in Table~\ref{t:out-of-sample}.
    \label{f:FFTM_vs_CAMB}
  }
\end{figure}

Next, we consider the range of scales over which we emulate the neutrino power spectra.  We set the upper limit of our range of wave numbers to $1~h/$Mpc, since beyond that we expect non-perturbative effects, such as the capture of neutrinos by halos, to dominate the power spectrum.  Meanwhile, Fig.~\ref{f:FFTM_vs_CAMB} demonstrates that our minimum wave number should not be much larger than $\sim 0.001~h/$Mpc.  Below that value, all power spectra shown agree with \cambcode{} to $\leq 5\%$, with percent-level agreement for nearly all models below $k=0.002~h/$Mpc.  Above $k=0.001~h/$Mpc, fractional differences rise rapidly, reaching $\sim 20\%$ by $k=0.01~h/$Mpc.  Thus we choose to emulate $\log(\DDnu/\DDehnu)$ and $\log(\DDnuL/\DDehL)$ over the range $0.001~h/$Mpc~$\leq k \leq 1~h/$Mpc.

%-------------------------------------------------------------------------------
\subsection{Non-linear enhancement: A first look}
\label{subsec:tra:non-linear_enhancement_a_first_look}

The chief goal of this study is to quantify the non-linear clustering of neutrinos across a wide range of parameters. One tool we will use for this is the non-linear enhancement ratio, that is, the ratio of the \FFTM{} power spectrum to that from the Multi-Fluid Linear Response code \MFLR{} of \citet{Chen:2020bdf}:\footnote{\MFLR{} is publicly available at {\tt github.com/upadhye/MuFLR}~.}
\begin{equation}
  \Rnu(k,z)
  =
  {\DDnu}_{,\FFTM{}}(k,z) / {\DDnu}_{,\MFLR{}}(k,z).
  \label{e:def_Rnu}
\end{equation}
\MFLR{} allows the CDM+baryon fluid to cluster non-linearly while limiting neutrino clustering to the linear terms.  That is, the mode-coupling integrals $\Aa{acd,bef}{}$, hence also the bispectrum integrals $\Ia{acd,bef}{}$ and the non-linear correlations $\xi_{\alpha\ell}$, are set to zero, leaving only Eqs.~(\ref{e:eom_delta},\ref{e:eom_theta},\ref{e:eom_Phi}) to be solved for the neutrinos.

The enhancement ratio is instructive because it allows us to isolate directly the effects of non-linearity in the neutrino sector.  High-quality simulations using neutrino linear response include \citet{Ali-Haimoud:2012fzp}, \citet{McCarthy:2016mry}, and \citet{Liu:2017now}.  $\Rnu(k,z)$ quantifies the amount by which these underestimate neutrino clustering.  Furthermore, \citet{Chen:2022dsv,Chen:2022cgw} showed that \FFTM{} itself becomes inaccurate in the regime $\Rnu-1 \gtrsim 1$, indicating the scales on which a particle neutrino simulation is necessary for accurate predictions of neutrino clustering.  Section~\ref{sec:enh} will use emulation to isolate the impact upon $\Rnu$ of individual cosmological parameters.

\begin{figure}
  \includegraphics[width=86mm]{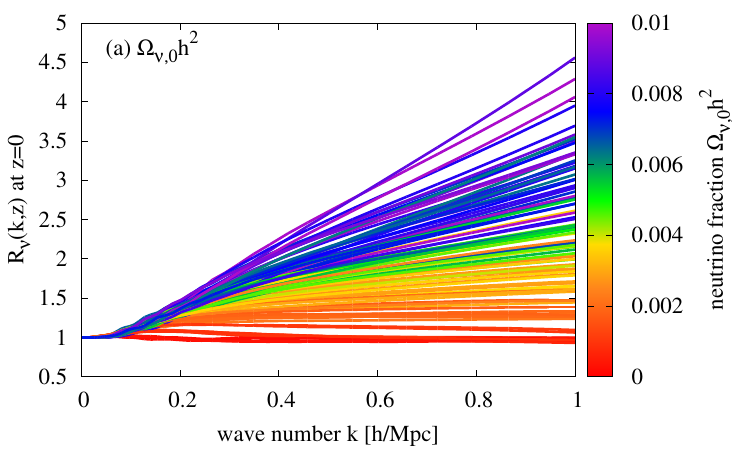}%

  \includegraphics[width=86mm]{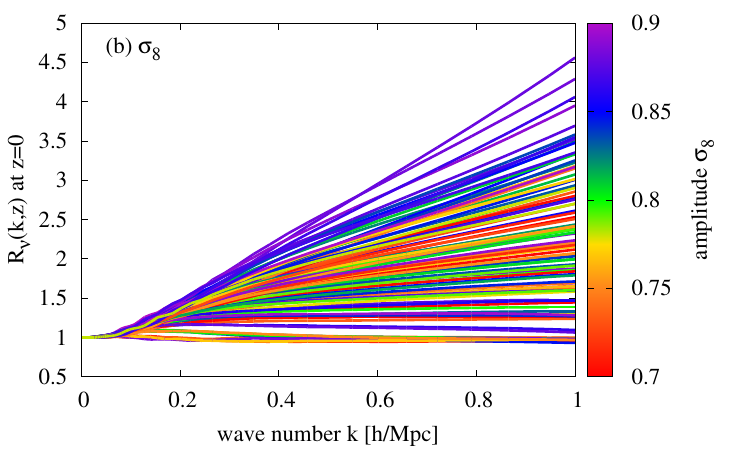}%
  
  \includegraphics[width=86mm]{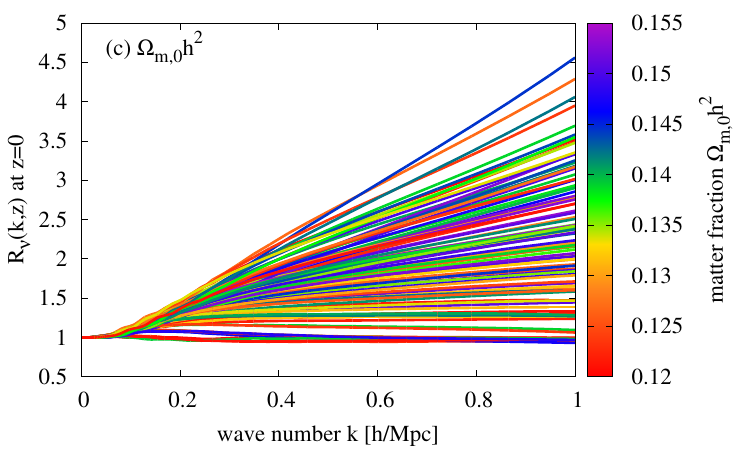}%

  \includegraphics[width=86mm]{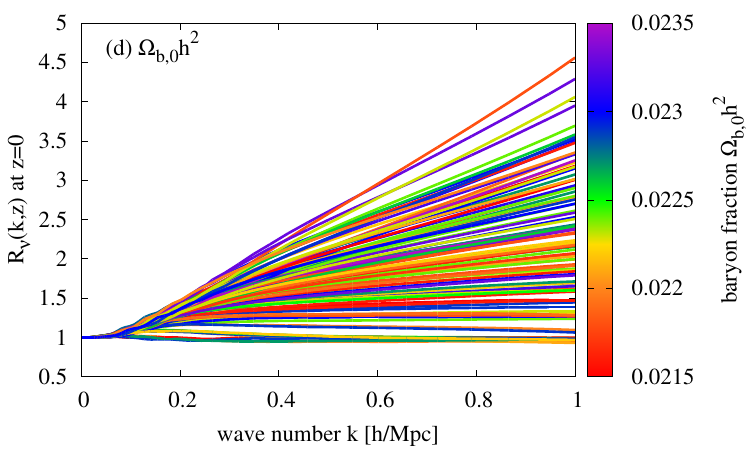}%
  \caption{
    Neutrino non-linear enhancement ratio of \EQ{e:def_Rnu} for each of the
    $\NM=101$ MT4 $\Mnu>0$ design points at redshift $z=0$.  Each curve is color-coded by (a)~$\Ono h^2$, (b)~$\sigma_8$, (c)~$\Omo h^2$, or (d)~$\Obo h^2$.
    \label{f:Rnu_first_look}
  }
\end{figure}

Figure~\ref{f:Rnu_first_look} provides a glimpse of the parameter-dependence of $\Rnu(k,0)$.  From Fig.~\ref{f:Rnu_first_look}~(a) we immediately see that $\Ono h^2 \propto \Mnu$ has a significant effect upon $\Rnu$.  Furthermore, non-linear corrections to the lightest neutrinos reduce $\Rnu$ slightly below unity.  This is not surprising, as non-linear corrections in the weakly non-linear regime are known to suppress even the CDM clustering~\citep{Bernardeau:2001qr}.    Evidently from Fig.~\ref{f:Rnu_first_look}, the clustering amplitude $\sigma_8$ also has a discernible effect upon $\Rnu$, while the impact of the total matter and baryon fractions are less obvious.  We will revisit the impact of different cosmological parameters upon $\Rnu(k,0)$ in Sec.~\ref{subsec:enh:parameter-sensitivity_of_Rnu}.

%%%%%%%%%%%%%%%%%%%%%%%%%%%%%%%%%%%%%%%%%%%%%%%%%%%%%%%%%%%%%%%%%%%%%%%%%%%%%%%%
\section{Emulation of the non-linear neutrino power spectrum}%%%%%%%%%%%%%%%%%%%
\label{sec:emu}%%%%%%%%%%%%%%%%%%%%%%%%%%%%%%%%%%%%%%%%%%%%%%%%%%%%%%%%%%%%%%%%%

%-------------------------------------------------------------------------------
\subsection{Emulation using \SEPIA{}}
\label{subsec:emu:emulation_using_sepia}

For each neutrino momentum decile $L$, we use the \SEPIA{} code of \citet{SEPIA} to determine the values of the principal component basis functions $\phi^{(L)}_j(k_i,z_i)$ as well as sample from the posteriors of the hyperparameters $\beta^{(L)}_{j\ell}$, $\lamu{j}^{(L)}$, and $\lamw{j}^{(L)}$.  We use $\NPC=50$ principal components for each $L$, comparable to the $45$ used for the MT4 emulator.  Hyperparameters are optimized in \SEPIA{} using $\Nmc=50000$ Markov chain Monte Carlo steps.  Since the error on the total neutrino power spectrum for $\Nmc$ of $10000$, $20000$, and $50000$ is, respectively, $3.92\%$, $3.86\%$, and $3.48\%$, measured against ten randomly-chosen models outside of the MT4 sample, we regard $\Nmc=50000$ to have converged.

\SEPIA{} is designed to draw weights from the probability distribution of \EQ{e:hat_p}.  We instead prefer a deterministic emulator such as that described in \citet{Heitmann:2009cu}, which uses the mean weights $\bar W_j^{(L)}(\vec C\,)$ of \EQ{e:bar_Wj} as the emulator prediction.  These depend on the hyperparameters' posterior means $\hat\beta^{(L)}_{j\ell}$, $\lamuh{j}^{(L)}$, and $\lamwh{j}^{(L)}$.  Appendix~\ref{app:imp} shows how to obtain these quantities from \SEPIA{}.

Our goal is now in sight: the mean weight $\bar W^{(L)}_j(\vec C\,)$ of \EQ{e:bar_Wj}, for each decile $L$ and principal component $j$, for a given cosmological parameter vector $\vec C$.  The final ingredient needed to emulate $\bar W^{(L)}_j(\vec C\,)$ is the so-called Kriging basis $[\hat\Rscr_j^{*(L)}]^{-1} \vec w_j^{\,*(L)}$.  With fixed $L$ and $j$, the quantity $\hat\Rscr_j^{*(L)}$ is an $\NM\times\NM$ matrix, $[\hat\Rscr_j^{*(L)}]^{-1}$ is its matrix inverse, and $\vec w_j^{\,*(L)}$ and $[\hat\Rscr_j^{*(L)}]^{-1} \vec w_j^{\,*(L)}$ are vectors of length $\NM$.  Since $\hat\Rscr_j^{*(L)}$ and $\vec w_j^{\,*(L)}$ are both known, we find the Kriging basis by solving the linear system:

\begin{equation}
  %\sum_{m=0}^{\NM-1} \hat\Rscr^{(L)}_{jnm} X^{(L)}_{jm} = w^{*(L)}_{jn}
  \hat\Rscr_j^{*(L)} X^{(L)}_{j} = \vec w_j^{\,*(L)}
\end{equation}
for $X^{(L)}_{j}$, which equals $[\hat\Rscr_j^{*(L)}]^{-1} \vec w_j^{\,*(L)}$.

Since the Kriging basis is independent of $\vec C$, we compute it once and save the result.  Now, given $\vec C$, we may readily compute $r_{jm}^{\:*}(\vec C\,)$ of \EQ{e:def_rjm} for each $m=0,\ldots,\NM-1$.  The dot product of the vector $\vec r_j^{\:*}(\vec C\,)$ with the Kriging basis gives $\bar W^{(L)}_j(\vec C\,)$, as in \EQ{e:bar_Wj}.  The emulated reduced neutrino power $\Pscr^{(L)}(k_i,z_i,\vec C) = \log(\DDnuL/\DDehL)$ for decile $L$ is given by:

\begin{equation}
  \Pscr^{(L)}(k_i,z_i,\vec C)
  =
  \mu^{*(L)}_i
  + \sigma^{*(L)} \sum_{j=0}^{\NPC-1} \bar W^{(L)}_j(\vec C\,) \phi^{(L)}_j(k_i,z_i),
\end{equation}
and the total neutrino power spectrum $\DDnu(k_i,z_i)$ is formed by averaging over the individual decile powers $\DDnuL(k_i,z_i)$:
\begin{equation}
\DDnu(k_i,z_i) = \left[ \frac{1}{10} \sum_{L=0}^9 \sqrt{\DDnuL(k_i,z_i)} \right]^2.
\label{e:decile_average}
\end{equation}
As an alternative, we could have emulated $\DDnu(k_i,z_i)$ separately.  However, we find averaging over decile powers to be more accurate.  Further, averaging ensures consistency between $\DDnu(k_i,z_i)$ and the individual $\DDnuL(k_i,z_i)$, which should obey \EQ{e:decile_average}.

%-------------------------------------------------------------------------------
\subsection{Tests of \cosmicenu{}}
\label{subsec:emu:tests_of_cosmicenu}

\begin{table*}
  \caption{
    Cosmological parameters for the ten out-of-sample test models.  Each is a
    spatially-flat $\nu w$CDM model with $w(a) = w_0 + (1-a)w_a$.
    \label{t:out-of-sample}
  }
  \begin{center}
    \begin{tabular}{l||cccccccc}
      model
      & $\Omo h^2$ & $\Obo h^2$ & $\Ono h^2$ & $\sigma_8$
      & $h$        & $n_{\rm s}$ & $w_0$      & $w_a$      \\
      \hline
      E001
      & $0.1433 $ & $0.02228 $ & $ 0.008078$ & $0.8389 $
      & $0.7822 $ & $0.9667 $ & $ -0.8000$ & $ -0.0111$ \\
      E002
      & $0.1333 $ & $0.02170$ & $0.005311 $ & $0.8233 $
      & $0.7444 $ & $0.9778 $ & $-1.1560 $ & $-1.1220 $ \\
      E003
      & $0.1450 $ & $0.02184 $ & $0.003467 $ & $0.8078 $
      & $0.6689 $ & $0.9000 $ & $-0.9333 $ & $-0.5667 $ \\
      E004
      & $0.1367 $ & $0.02271 $ & $0.002544 $ & $0.8544 $
      & $0.8200 $ & $0.9444 $ & $-0.8889 $ & $-1.4000 $ \\
      E005
      & $0.1400 $ & $0.02257 $ & $0.009000 $ & $0.7300 $
      & $0.7067 $ & $0.9889 $ & $-0.9778 $ & $-0.8444 $ \\
      E006
      & $0.1350 $ & $0.02213 $ & $0.000700 $ & $0.8700 $
      & $0.7633 $ & $0.9111 $ & $-1.0220 $ & $0.5444 $ \\
      E007
      & $0.1383 $ & $0.02199 $ & $0.007156 $ & $0.7456 $
      & $0.6500 $ & $0.9556 $ & $-1.1110 $ & $1.1000 $ \\
      E008
      & $0.1300 $ & $0.02286 $ & $0.006233 $ & $0.7922 $
      & $0.8011 $ & $1.0000 $ & $-1.0670 $ & $0.2667 $ \\
      E009
      & $0.1417 $ & $0.02300 $ & $0.004389 $ & $0.7767 $
      & $0.7256 $ & $0.9222 $ & $-0.8444 $ & $0.8222 $ \\
      E010
      & $0.1317 $ & $0.02242 $ & $0.001622 $ & $0.7611 $
      & $0.6878 $ & $0.9333 $ & $-1.2000 $ & $-0.2889 $ \\
    \end{tabular}
  \end{center}
\end{table*}

Now that our \cosmicenu{} emulator is complete, we quantify its accuracy.  We first consider the total neutrino power spectrum $\DDnu(k,z)$, the main goal of this study.  Power spectra of individual momentum deciles, $\DDnuL(k,z)$, are covered at the end of this subsection.

We begin with the most accurate error measurement, which compares emulator predictions to \FFTM{} computations for a set of ten test models outside of the MT4 design set.  Table~\ref{t:out-of-sample} lists the cosmological parameters of these out-of-sample models, E001 through E010.  They cover a large range of $\Ono h^2$ from $0.0007$, for E006, to $0.009$, for E005, and allow for a substantial variation in the dark energy equation of state.

\begin{figure}
  \includegraphics[width=89mm]{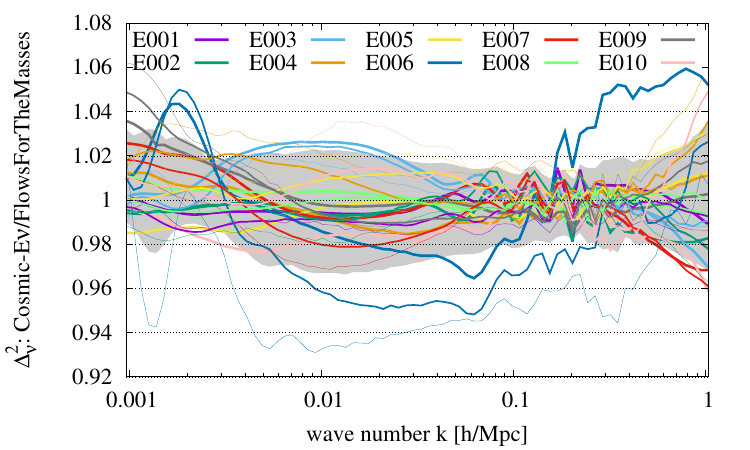}%
  \caption{
    \cosmicenu{} vs.~\FFTM{} for the total neutrino power spectra
    $\DDnu(k,z)$ of the out-of-sample models of Table~\ref{t:out-of-sample}. For each model, lines show the redshifts $3$, $1$, and $0$, in order of increasing thickness.  The gray shaded region is the mean plus-or-minus one standard deviation, maximized over all emulated redshifts.  Across all $k$ and $z$, this error is less than $3.5\%$.
    \label{f:D2nu_out-of-sample}
  }
\end{figure}

Figure~\ref{f:D2nu_out-of-sample} compares the total neutrino power spectra of \cosmicenu{} and \FFTM{} for the ten out-of-sample models of Table~\ref{t:out-of-sample}.   For each of the $27$ redshifts emulated and at each $k$, the mean and standard deviation of the \cosmicenu{}-to-\FFTM{} ratio are computed.  The $1\sigma$ error interval is the region within one standard deviation of the mean ratio at each $k$ and $z$ emulated, {\em i.e.}, the standard deviation across cosmologies for a single redshift and wave number.  At each $k$, the gray shaded region shows the $1\sigma$ error interval maximized over redshift.  The $1\sigma$ error, the difference between this region and unity, is everywhere less than $3.5\%$.  The maximum error across all ten models, over all $k$ and $z$, is $6.9\%$.  The largest error is for model E006, which has a  small neutrino density.

\begin{figure}
  \includegraphics[width=87mm]{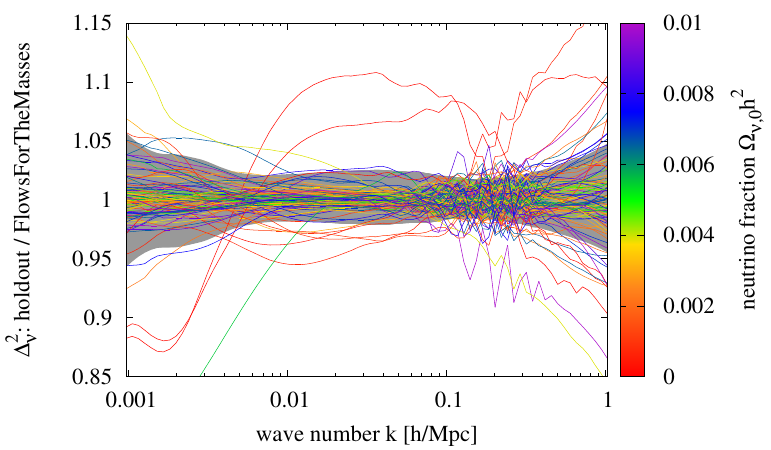}%
  \caption{
    Leave-one-out holdout tests of \cosmicenu{}.  Lines show the holdout-to-\FFTM{} power spectrum ratios at $z=0$, with colors corresponding to $\Ono h^2$.  The dark gray shaded region is the mean plus-or-minus one standard deviation, maximized over all emulated redshifts; the maximum $1\sigma$ error across all $k$ and $z$ is $5.7\%$.
    \label{f:D2nu_holdout}
  }
\end{figure}

Another emulator error estimate, the leave-one-out holdout test, is performed directly through \SEPIA{}.  For each model $m$ of the $\NM$ models in the training set, \SEPIA{} builds an emulator with all $\NM-1$ models excluding $m$, then compares the resulting emulator prediction $\Pscr_{im}$ at $\vec C^*_m$ with the excluded training data $\Pscr^*_{im}$ for that model.  Since leaving out model $m$ creates a gap in the training data at $\vec C^*_m$, holdout tests tend to overestimate the emulator error, particularly in the context of a space-filling design.

Figure~\ref{f:D2nu_holdout} shows the results of leave-one-out holdout tests of \cosmicenu{}.  The maximum $1\sigma$ error is now $5.7\%$, about $60\%$ higher than the out-of-sample test.  Nevertheless, at all but the largest scales, $k < 0.002~h/$Mpc, and the smallest scales, $k > 0.8~h/$Mpc, the error is $4\%$ or less.  Meanwhile, the largest holdout test error across all $k$ and $z$, and all $101$ models, is $25.4\%$.

\begin{figure}
    \includegraphics[width=89mm]{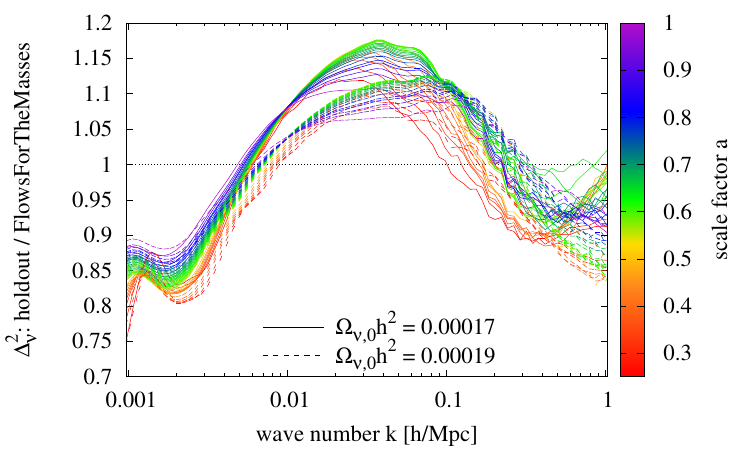}%
    \caption{
      Holdout tests for the two lowest-neutrino-mass models, which also have the largest errors in Fig.~\ref{f:D2nu_holdout}, color-coded by the scale factor $a$.  Solid (dashed) curves correspond to $\Ono h^2 = 0.00017$ ($\Ono h^2 = 0.00019$).
    \label{f:holdout_M017_M066}
    }
\end{figure}

The two models with the largest holdout test errors are further studied in Fig.~\ref{f:holdout_M017_M066}.  Out of the models in the MT4 training set, these two have the smallest $\Ono h^2$; they are the only models allowing $\Ono h^2$ to fall below half of its lower bound $\approx 0.00064$ from neutrino oscillation experiments.  Evidently, the $25\%$ errors noted above are due to sharp falls in the \cosmicenu{} predictions at the earliest times and largest scales.  Errors are under $20\%$ for all $k \geq 0.0012~h/$Mpc and all $z$.  Since the holdout test errors of Fig.~\ref{f:D2nu_holdout} overestimate the more accurate out-of-sample errors of Fig.~\ref{f:D2nu_out-of-sample} by about $60\%$, we estimate \cosmicenu{} errors of $10\%-12\%$  at the lowest neutrino masses, in the range $k \geq 0.0012~h/$Mpc.

\begin{figure*}
  \begin{center}
  \includegraphics[width=180mm]{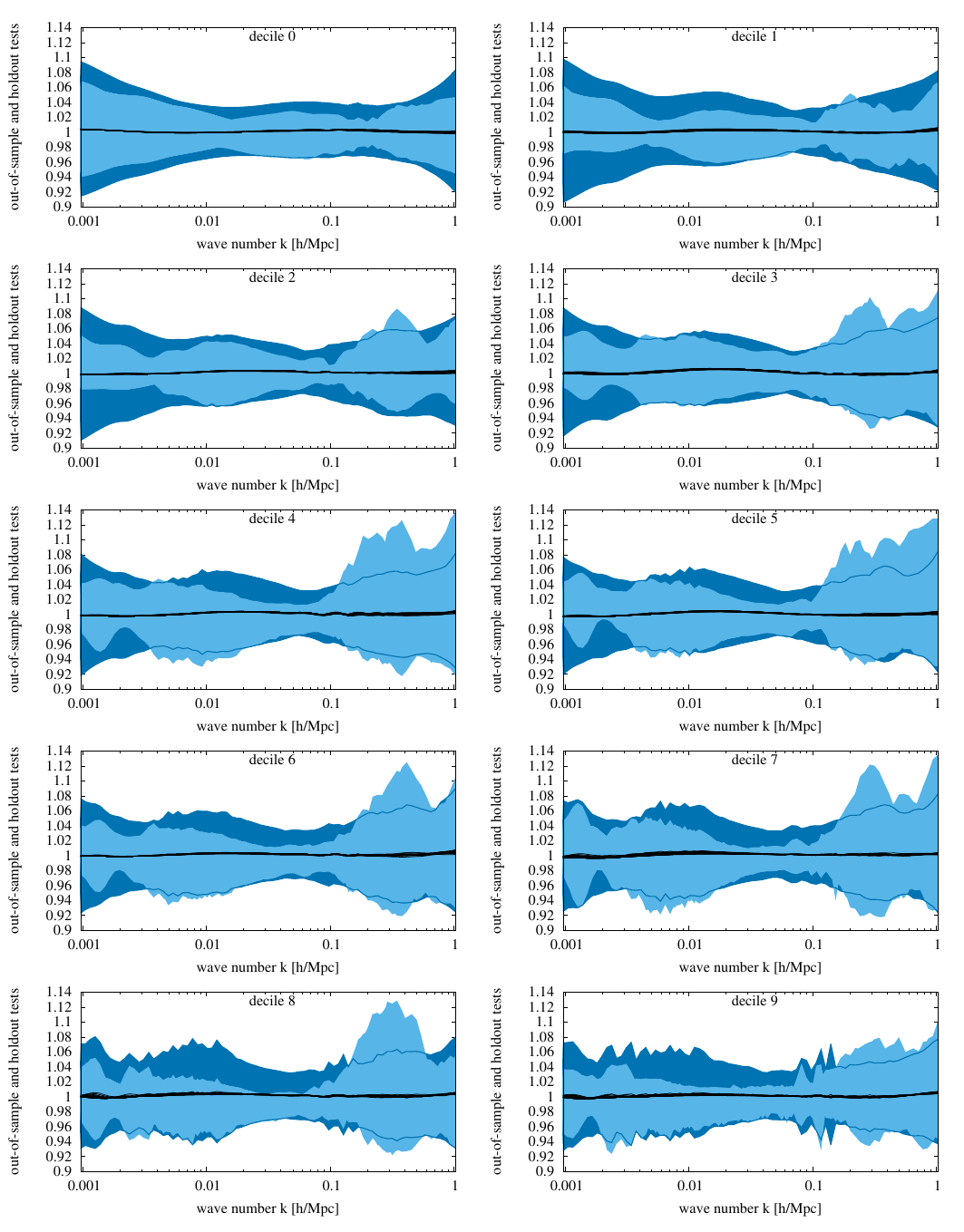}%
  \end{center}
  \caption{
    Out-of-sample (light shaded regions) and holdout (dark shaded regions) tests for the individual neutrino decile power spectra $\DDnuL(k,z)$.  Shaded regions show the mean plus-or-minus one standard deviation, maximized over all emulated redshifts.
    \label{f:D2nuL_outofsample_holdout}
  }
\end{figure*}

Finally, we compare individual neutrino momentum deciles between \cosmicenu{} and \FFTM{}.  In multi-fluid perturbation theories such as ours of Sec.~\ref{subsec:bkg:nonlinear_perturbation_theory_for_neutrinos}, the neutrino flow speed $\va$ behaves similarly to a sound speed, and the resulting power spectrum exhibits oscillatory behavior on sufficiently small scales.  Averaging over a large number of flows eliminates these oscillations.  However, each decile averages over only $5$ flows, rather than $50$ for the total neutrino power, making the small-scale decile powers noisier and more difficult to emulate.  This is especially true for the higher-momentum deciles for which individual-flow oscillations are more prominent.  Thus we expect less accuracy in $\DDnuL(k,z)$, particularly for high $L$, than in the $\DDnu(k,z)$ shown above.

Figure~\ref{f:D2nuL_outofsample_holdout} combines out-of-sample tests and leave-one-out holdout tests for each of the ten momentum deciles.  The $1\sigma$ errors shown are in line with our expectations above.  For deciles $L=0$ and $1$, errors from out-of-sample tests are $< 7\%$, or about twice the maximum error on $\DDnu(k,z)$.  For higher $L$, errors remain at this level for $k \leq 0.1~h/$Mpc but grow substantially at large $k$, rising to nearly four times the $\DDnu(k,z)$ emulator error at $k=1~h/$Mpc.

%%%%%%%%%%%%%%%%%%%%%%%%%%%%%%%%%%%%%%%%%%%%%%%%%%%%%%%%%%%%%%%%%%%%%%%%%%%%%%%%
\section{Comparison to N-body simulations}%%%%%%%%%%%%%%%%%%%%%%%%%%%%%%%%%%%%%%
\label{sec:nbd}%%%%%%%%%%%%%%%%%%%%%%%%%%%%%%%%%%%%%%%%%%%%%%%%%%%%%%%%%%%%%%%%%

%-------------------------------------------------------------------------------
\subsection{Variation of the $\nu$ implementation}
\label{subsec:nbd:variation_of_the_nu_implementation}

The previous section quantified the precision with which \cosmicenu{} reproduced its underlying \FFTM{} perturbation theory.  Next, we consider its accuracy relative to numerical simulations of the massive neutrino power spectrum.  \citet{Euclid:2022qde} carried out an extensive comparison of neutrino simulation methods for a spatially-flat $\nu\Lambda$CDM model with $\Mnu=0.15$~eV ($\Ono h^2 = 0.00161$), $\Omo h^2 = 0.1432$, $\Obo h^2 = 0.022$, $A_{\rm s} = 2.215\times 10^{-9}$ ($\sigma_8=0.815$), $h=0.67$, and $n_{\rm s} = 0.9619$.  Though all of the neutrino power spectra agree at about the percent level in the linear regime, $k \lesssim 0.1~h/$Mpc, the different methods are discrepant at the $30\%-40\%$ level by $k=1~h/$Mpc.  

Even within a given simulation method, choices of initial conditions and simulation parameters have a substantial impact on the massive neutrino power spectrum.  \citet{Sullivan:2023ntz} implemented the "tiling" method of \citet{Banerjee:2018bxy}.  They find that the number of allowed neutrino momentum directions is the most important parameter for determining convergence, and they judge their highest-direction-number run to have converged at the $\sim 10\%$ level up to half of the neutrino Nyquist frequency, or $\sim 1~h/$Mpc in their standard run.  Combining this with the discrepancies among the different simulation methods, we may regard simulation uncertainty in the neutrino power at $k\lesssim 1~h/$Mpc to be in the $30\%-50\%$ range.

\begin{figure}
  \includegraphics[width=89mm]{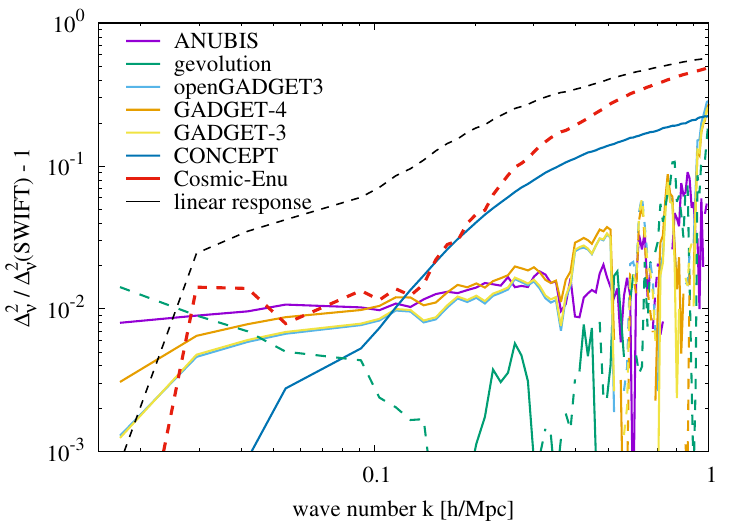}%
  \caption{
    Comparison of \cosmicenu{} and \MFLR{} (linear response) to N-body neutrino
    power spectra, for $\Mnu=0.15$~eV, at $z=0$, computed using a variety of
    methods by \citet{Euclid:2022qde}.  Solid (dashed) lines represent power spectra that are greater (less) than that of {\tt SWIFT}.  Ratios have been smoothed using a centered $10$-point moving average.
    \label{f:codecomp_methods}
  }
\end{figure}

Figure~\ref{f:codecomp_methods} compares \cosmicenu{} and \MFLR{} linear response to the results of \citet{Euclid:2022qde}.  Each power spectrum is compared to that computed using the {\tt SWIFT} simulation of \citet{SWIFT:2023dix}, with their ratio smoothed using a centered $10$-point moving average.   Compared with {\tt SWIFT}, the \concept{} code of \citet{Dakin:2017idt} predicts $22\%$ more power at $k=1~h/$Mpc, while {\tt gevolution}~\citep{Adamek:2016zes} predicts $\approx 15\%-20\%$ less.

\cosmicenu{} agrees closely with all of the simulated $\DDnu$ at low $k$.  Its predicted power spectrum falls below that of  {\tt SWIFT} by $\leq 3\%$ up to $k=0.17~h/$Mpc, or three times the free-streaming wave number for this neutrino mass.  This power deficit grows rapidly, rising to $19\%$ at $k=0.4~h/$Mpc and $49\%$ at $k=1~h/$Mpc, somewhat greater than, but comparable to, the scatter between different N-body simulation methods.  Thus \FFTM{}  and the \cosmicenu{} emulator appear to provide accurate computations of the non-linear neutrino power, given the current level of simulation uncertainty.

Most of the scatter among the N-body methods at wave numbers $k \approx 0.5~h/$Mpc is due to a systematic difference between particle-based methods, such as {\tt SWIFT}, and \concept{}, which integrates the massive neutrino fluid equations on a grid.  {\em A priori} we have no reason to consider one of these more accurate.  However, if we exclude \concept{} as an outlier among the N-body simulations, then the range spanned by the remaining simulations  drops significantly, and the \cosmicenu{} power deficit exceeds this range by a factor of a few for $k \gtrsim 0.5~h/$Mpc.  Thus it is not clear whether the \cosmicenu{} small-scale power deficit is due to systematic uncertainties among the different non-linear methods or to a genuine non-perturbative effect such as the capture of neutrinos by CDM+baryon halos.

Also shown in Fig.~\ref{f:codecomp_methods} is the \MFLR{} linear response power spectrum.  As with \cosmicenu{}, its power is less than that of the {\tt SWIFT} simulation used as a reference.  Its power deficits relative to {\tt SWIFT} are significantly larger than those of \cosmicenu{}, $6.1$ times larger at $k=0.1~h/$Mpc and $1.8$ times larger at $k=0.4~h/$Mpc.  Thus \cosmicenu{} represents a significant accuracy improvement over linear response approximations.

%-------------------------------------------------------------------------------
\subsection{Variation of the $\nu$ mass}
\label{subsec:nbd:variation_of_the_nu_mass}

\begin{figure}
  \includegraphics[width=89mm]{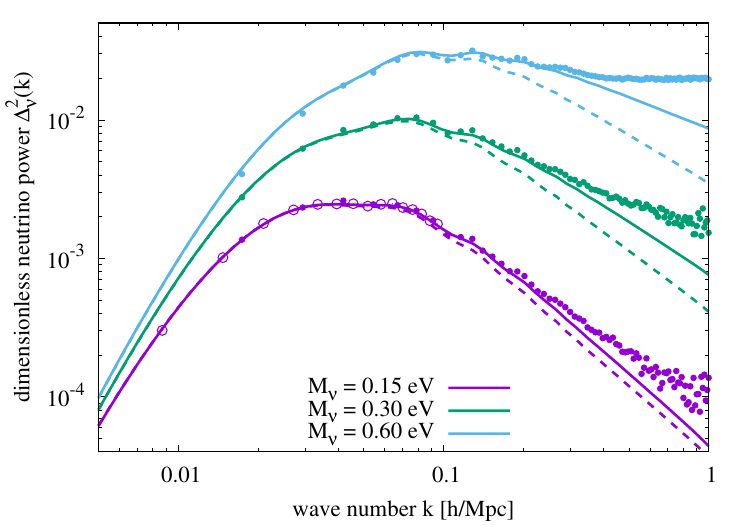}%
  
  \includegraphics[width=89mm]{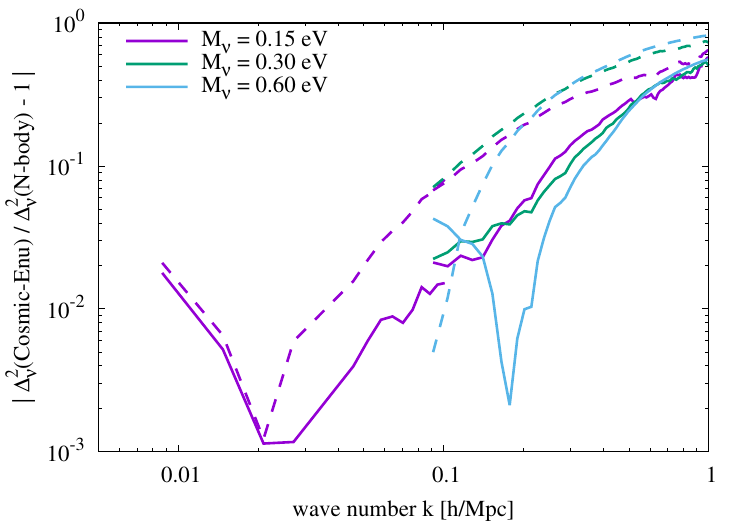}%
  \caption{
    Comparison of \cosmicenu{} (solid) and \MFLR{} (dashed)
    to {\tt GADGET-}3 N-body neutrino power spectra of
    \citet{Euclid:2022qde}, at $z=0$, for neutrino masses $\Mnu$ ranging
    from $0.15$~eV to $0.6$~eV.  For $\Mnu=0.15$~eV, the high-resolution
    $1024^3$-particle simulations of that reference are used.
    (Top)~N-body power spectra are shown as filled circles for
    $(512~{\rm Mpc})^3$ simulation volumes and open circles for the
    $(1024~{\rm Mpc})^3$ $\Mnu=0.15$~eV simulation.
    (Bottom)~Fractional errors in \cosmicenu{} and \MFLR{} compared with the
    $(512~{\rm Mpc})^3$-box simulations at $k \geq 0.1~h/$Mpc and the
    $(1024~{\rm Mpc})^3$-box simulation below that wave number.
    \label{f:codecomp_Mnu}
  }
\end{figure}

Next, we consider variations in $\Mnu$.  \citet{Euclid:2022qde} varied $\Mnu$ from $0.15$~eV to $0.6$~eV, in their flat $\nu\Lambda$CDM model with $\Omo h^2$, $\Obo h^2$, $h$, $A_{\rm s}$, and $n_{\rm s}$ fixed to the values of Sec.~\ref{subsec:nbd:variation_of_the_nu_implementation}.  Fixing the initial power spectrum amplitude  $A_{\rm s} = 2.215\times 10^{-9}$ implies $\sigma_8=0.815$ for $\Mnu=0.15$~eV, $\sigma_8=0.776$ for $\Mnu=0.3$~eV, and $\sigma_8=0.731$ for $\Mnu=0.6$~eV.  Their simulations for $\Mnu=0.3$~eV and $0.6$~eV use $512^3$ neutrino particles in a $(512~{\rm Mpc})^3$ box.

Figure~\ref{f:codecomp_Mnu} compares \cosmicenu{} and \MFLR{} to the \citet{Euclid:2022qde} simulations with varying $\Mnu$.  For $k \sim 0.1~h/$Mpc, \cosmicenu{} agrees with the simulations at the $2\%-4\%$ level, while \MFLR{} underpredicts power by $7\%$ for the lower masses.  At larger $k$, both errors quickly increase.  \cosmicenu{} errors at $k=0.4~h/$Mpc are $17\%$ and $14\%$ for $\Mnu=0.3$~eV and $0.6$~eV, respectively, while those for \MFLR{} are respectively $46\%$ and $49\%$.  Above $k=0.4~h/$Mpc, particularly for $\Mnu=0.6$~eV, the slopes of the N-body power spectra in Fig.~\ref{f:codecomp_Mnu}~(Top) flatten in a manner not captured by either \cosmicenu{} or \MFLR{}.

%-------------------------------------------------------------------------------
\subsection{Rapidly-evolving dark energy}
\label{subsec:nbd:rapidly-evolving_dark_energy}

\begin{figure}
  \includegraphics[width=89mm]{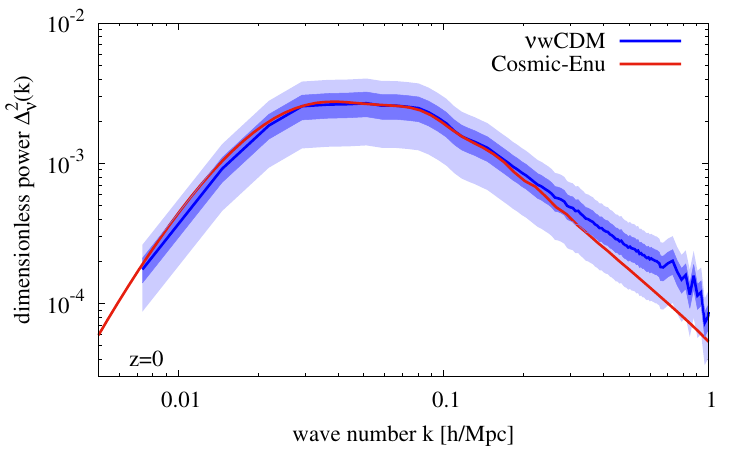}%

  \includegraphics[width=89mm]{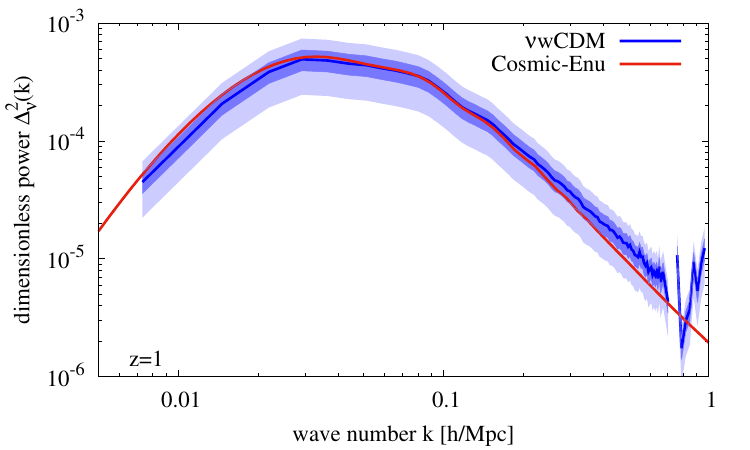}%
  \caption{
    Test of \cosmicenu{} for a model with rapidly-evolving dark energy,
    $w_0=-1.036$ and $w_a=-0.872$, with $\Ono h^2 = 0.0015$ ($\Mnu=0.14$~eV),
    at (Top)~$z=0$ and (Bottom)~$z=1$.  Our \bahamasxl{} N-body simulation
    is shown for comparison. Inner and outer shaded bands around
    the simulated power spectra show regions within $20\%$ and $50\%$, respectively.
    \label{f:bahamasXL_DE}
  }
\end{figure}

Since neutrino mass bounds are dependent upon constraints on the growth factor of large-scale structure, they are degenerate with variations in the dark energy equation of state.  For example, \citet{Upadhye:2017hdl} found a factor-of-three degradation in the $95\%$-confidence $\Mnu$ bound when $w_0$ and $w_a$ were allowed to vary.  In recognition of this degeneracy, MT4 and \cosmicenu{} allow for substantial variations in $w_0$ and $w_0+w_a$.  Here we test the accuracy of \cosmicenu{} for such a rapidly-varying equation of state by comparison to a \bahamasxl{} N-body simulation.

Our \bahamasxl{} simulation is a part of a forthcoming suite of simulations designed to investigate the simultaneous variation of parameters describing physics beyond the $\Lambda$CDM model. Its cosmological parameters are $\Omo h^2=0.1429476$, $\Obo h^2=0.0240724$, $\Ono h^2=0.0015$, $\sigma_8=0.841333$, $h=0.613333$, $n_s=0.9544$, $w_0=-1.036$, and $w_a=-0.872$.  It tracked $N=1260^3$ particles in a cubic volume with box length $1400$~Mpc and periodic boundary conditions. Its 3LPT initial conditions were produced at $z=31$ using {\tt Monofonic}~\citep{Rampf:2020ety,Hahn:2020lvr}, and it tracked massive neutrinos using  the $\delta f$ method of~\citet{Elbers:2020lbn,Elbers:2022tvb}.  The simulation was evolved to $z=0$ using {\tt SWIFT}~\citep{Schaller:2023hzn}.

Figure~\ref{f:bahamasXL_DE} tests \cosmicenu{} for this \bahamasxl{} simulation, with a rapidly-varying equation of state, at redshifts $0$ and $1$.  Inner and outer shaded bands show the regions within $20\%$ and $50\%$ of the N-body power, respectively.  The accuracy of \cosmicenu{} is in line with our previous $\nu\Lambda$CDM comparisons to the {\tt SWIFT} simulations: $20\%$ up to $k = 0.3~h/{\rm Mpc} - 0.4~h/{\rm Mpc}$, and $50\%$ up to $k \approx 1~h/$Mpc, with slightly higher accuracy at $z=1$.  Thus we conclude that even $|w_a| \sim 1$ does not diminish the accuracy of \cosmicenu{}.

This Section has quantified the accuracy of the \cosmicenu{} $\DDnu(k,z)$ emulator across a wide range of $\Mnu$, for a cosmological constant as well as a rapidly-evolving equation of state, by comparison to N-body simulations using a few very different massive neutrino simulation methods.  Its error at $z=0$ is fairly consistent across a wide range of methods: a few percent up to $k\approx 0.15~h/$Mpc, an $\approx 20\%$ power underestimate at $k=0.4~h/$Mpc, and an $\approx 50\%$ underestimate at $k=1~h/$Mpc.  Since $\DDnu(k,z)$ for $k \gg \kfs$ scales approximately as $\Mnu^4$~\citep{Ringwald:2004np,Wong:2008ws}, these underestimates at $k=0.4~h/$Mpc and $k=1~h/$Mpc are consistent with $5\%$ and $13\%$ biases in $\Mnu$, respectively.  

Before proceeding, we comment upon the discrepancy between the $14\%$ error in \FFTM{} reported in \citet{Chen:2022cgw}, over the entire range $k\leq 1~h/$Mpc, and the larger differences with N-body simulations evident in Figs.~\ref{f:codecomp_methods},~\ref{f:codecomp_Mnu}, and \ref{f:bahamasXL_DE} for $k \approx 1~h/$Mpc.  There are two possibilities: errors in the hybrid simulations of \citet{Chen:2022dsv} used to test \FFTM{}, and a systematic error causing the discrepancies between differing N-body implementations of neutrinos.  

Errors in \citet{Chen:2022dsv} may be due to a finite number of neutrino flows, residual shot noise, and a finite simulation volume.  Finite-flow-number errors in the linear response calculations of \citet{Chen:2020bdf} were found to be $\approx 10\%$, and non-linear neutrino clustering likely increases them somewhat.  Of course, increasing the number of flows, or sampling the Fermi-Dirac distribution more efficiently, will improve the accuracy of \FFTM{} as well as the simulations.  While the estimated simulation shot noise $(k L_{\rm sim})^3 / (2\pi^2 N_{\rm sim})$, for $N_{\rm sim}$ particles in a volume $L_{\rm sim}^3$, was subtracted from $\DDnu(k)$, residual shot noise remains.  In an effort to mitigate shot noise, \citet{Chen:2022dsv} chose a small box, $L_{\rm sim} = 128$~Mpc$/h$, at the cost of neglecting the contributions of larger modes to small-scale non-linear growth.

Meanwhile, the $30\%-40\%$ spread among the different simulations in Fig.~\ref{f:codecomp_methods} at $k=1~h/$Mpc suggests small-scale systematic errors in some of these methods.  The {\tt gevolution} power spectrum of \citet{Adamek:2016zes} is about $15\%-20\%$ lower than {\tt SWIFT} at $k=1~h/$Mpc, meaning that correcting the $14\%$ underestimate of \cosmicenu{} relative to \citet{Chen:2022dsv}, as well as the $\approx 10\%$ underestimate due to a finite number of flows, would put \cosmicenu{} within $5\%-10\%$ of {\tt gevolution}.  Moreover, in the $k\lesssim 0.2~h/$Mpc range where the N-body methods of Fig.~\ref{f:codecomp_methods} agree with one another to a few percent, \cosmicenu{} also agrees with them at that level.  The {\tt SWIFT}-\cosmicenu{} difference rises along with the \concept{}-{\tt gevolution} difference in Fig.~\ref{f:codecomp_methods}.  Thus we cannot conclusively attribute the discrepancy between the $14\%$ \FFTM{} error estimate of \citet{Chen:2022cgw} and the $49\%$ {\tt SWIFT}-\cosmicenu{} difference to errors in \citet{Chen:2022dsv}.

%%%%%%%%%%%%%%%%%%%%%%%%%%%%%%%%%%%%%%%%%%%%%%%%%%%%%%%%%%%%%%%%%%%%%%%%%%%%%%%%
\section{Non-linear enhancement}%%%%%%%%%%%%%%%%%%%%%%%%%%%%%%%%%%%%%%%%%%%%%%%%
\label{sec:enh}%%%%%%%%%%%%%%%%%%%%%%%%%%%%%%%%%%%%%%%%%%%%%%%%%%%%%%%%%%%%%%%%%

\subsection{Parameter-sensitivity of the enhancement ratio}
\label{subsec:enh:parameter-sensitivity_of_Rnu}

\begin{figure*}
  \includegraphics[width=180mm]{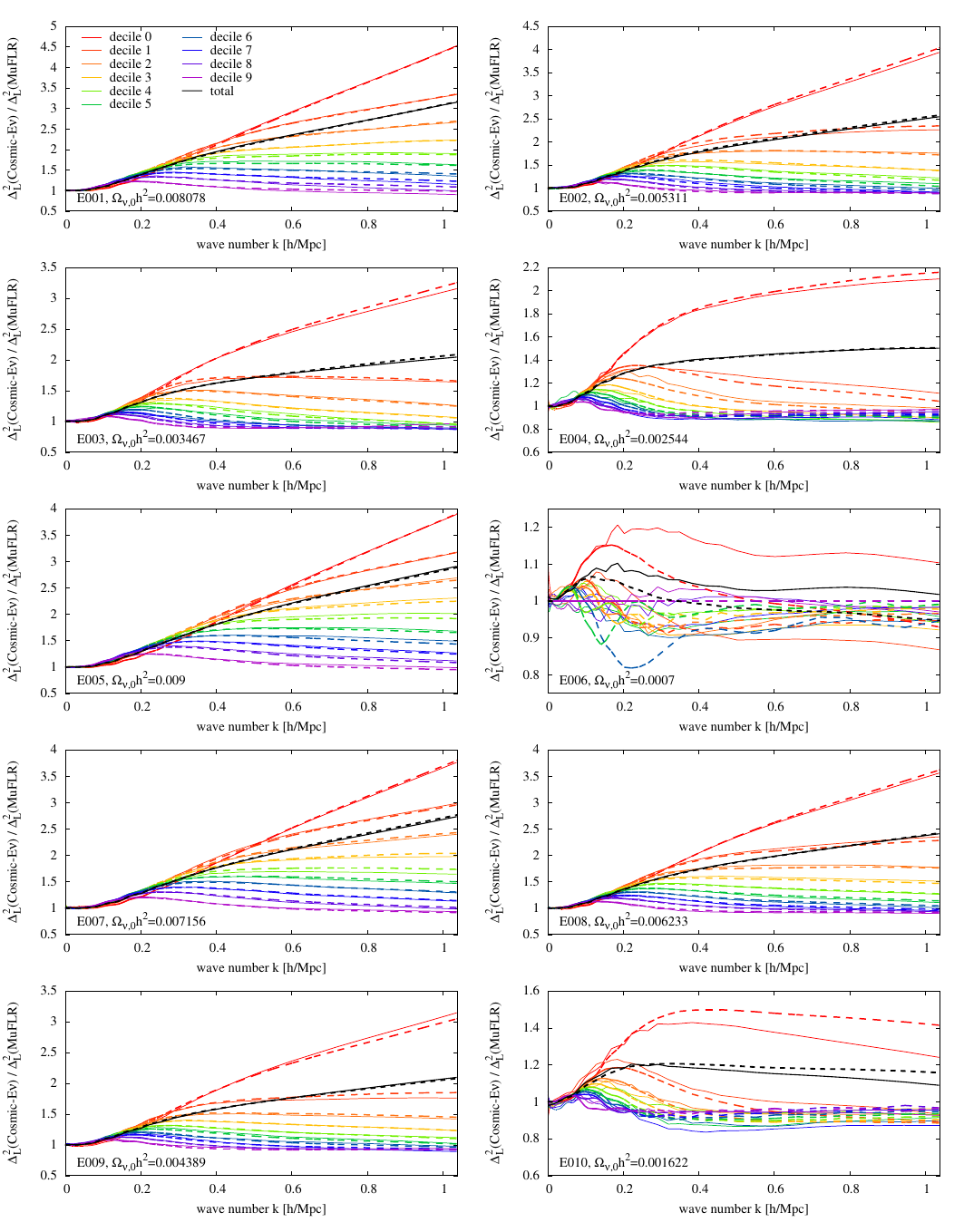}%
  \caption{
    $\Rnu(k,z)$ for each $L$, at $z=0$, for the out-of-sample cosmologies of Table~\ref{t:out-of-sample}, computed using \cosmicenu{} (solid) and \FFTM{} (dashed).
    \label{f:Rnu_out-of-sample}
  }
\end{figure*}

Now that we have quantified the accuracy of \cosmicenu{}, we may use it to study the non-linear clustering of massive neutrinos.  We focus here on the non-linear enhancement ratio $\Rnu(k,z)$ of \EQ{e:def_Rnu}, that is, the ratio of the neutrino power spectra using \FFTM{} and \MFLR{}, with the CDM+baryon treatment held fixed.  We emulate $\Rnu$ by taking the ratio of \cosmicenu{} to a \MFLR{} emulator.  

Figure~\ref{f:Rnu_out-of-sample} compares perturbative (dashed) and emulated (solid) calculations of $\Rnu$ at $z=0$ for the out-of-sample models of Table~\ref{t:out-of-sample}.  A couple trends are evident.  Firstly, the total neutrino power is typically more accurate at high $k$ than individual decile powers.  At $k=1~h/$Mpc, the emulated $\Rnu$ agrees with the \FFTM{} computation to better than $2\%$ for eight of the ten models.  For decile $0$, this error rises to $3.2\%$, and for decile $1$ to $4.8\%$.

Secondly, the lower deciles and higher neutrino masses tend to have smaller errors.  This is due to the fact that larger $L$ and smaller masses lead to larger average velocities, hence more prominent oscillatory behavior in the free-streaming limit, making these flows difficult to emulate.  Both of these trends are consistent with the individual-decile out-of-sample and holdout tests of Sec.~\ref{subsec:emu:tests_of_cosmicenu}.

\begin{figure}
  \includegraphics[width=89mm]{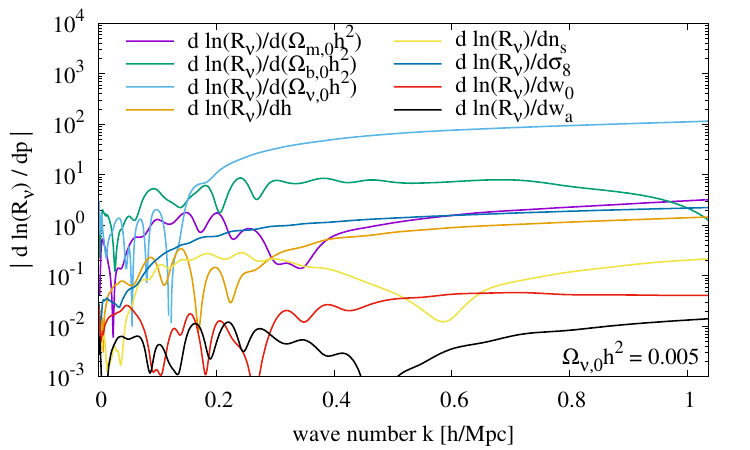}%
  \caption{
    Sensitivity of $\Rnu(k,z)$ at $z=0$ to variations of the eight
    cosmological parameters shown in Table~\ref{t:MT4_range}.
    \label{f:dlnR_dp}
    }
\end{figure}

Next, we consider the sensitivity of $\Rnu$ to the cosmological parameters.  As a fiducial model at which to test this sensitivity, we choose a $\Lambda$CDM model in which each parameter except for $w_0$ and $w_a$ is set to the midpoint of its range in Table~\ref{t:MT4_range}. Figure~\ref{f:dlnR_dp} shows the derivative of $\log\Rnu$ with respect to each parameter about this fiducial model.  We have checked that the results are qualitatively similar for $\Ono h^2 = 0.002$.

Above $k \approx \kfs = 0.16~h/$Mpc, the dominant effect is a rise in $\Rnu$ with $\Ono h^2 \propto \Mnu$.  This can be understood by noting that, with all parameters in Table~\ref{t:MT4_range} other than $\Ono h^2$ held fixed, the small-scale linear and linear-response neutrino power scales as $(\Ono h^2)^4$, while the non-linear power rises relative to the LR power.  We can estimate the non-linear enhancement to the linear scaling law from Fig.~\ref{f:dlnR_dp}:
\begin{equation}
  \frac{\partial \log(\DDnu[{\rm nonlin}])}{\partial \log(\Ono h^2)}
  =
  \frac{\partial \log(\DDnu[{\rm LR}])}{\partial \log(\Ono h^2)}
  +
  \frac{\partial \log(\Rnu)}{\partial \log(\Ono h^2)}.
  \label{e:mass-scaling_enhancement}
\end{equation}
The final term on the right, the logarithmic derivative of $\Rnu$, is $\Ono h^2$ times $d \ln(\Rnu)/d (\Ono h^2)$ from the figure, or about $0.57$ at $k=1~h/$Mpc~$\approx 6  \kfs$.  A similar analysis for $\Ono h^2 = 0.002$ finds this scaling enhancement to be $0.35$ at $k=0.4~h/$Mpc~$\approx 6 \kfs$ and $0.59$ at $k=1~h/$Mpc~$\approx 15 \kfs$, suggesting a rise in this scaling enhancement with both $\Ono h^2$ and $k/\kfs$.  Section~\ref{subsec:enh:relative_clustering_in_free-streaming_limit} will explore this enhancement further.

At $k \gtrsim \kfs=0.16~h/$Mpc, the next most significant parameter for determining $\Rnu$ is the physical baryon fraction $\Obo h^2$.  The oscillatory nature of $\partial \ln \Rnu / \partial (\Obo h^2)$ for $k\lesssim 0.8~h/$Mpc, and its decline for $k \gtrsim 0.9~h/$Mpc, suggests that $\Obo h^2$ affects $\Rnu$ primarily through the baryon acoustic oscillations (BAO).  Modifications to the BAO, in turn, are amplified by the non-linear clustering of neutrinos.

\begin{table}
    \caption{
      Parameter sensitivities from Fig.~\ref{f:dlnR_dp} multiplying the
      $95\%$~CL intervals $\Delta p$ for the $\nu w$CDM model of
      \citet{Upadhye:2017hdl}, analyzed using the combination of Planck, BOSS, 
      and JLA supernova data, marginalized over a five-parameter bias model.
      Third and fourth columns, respectively, use sensitivities at
      $k=0.4~h/$Mpc and $k=1~h/$Mpc. We assume
      $\Delta \Omo h^2 \approx \Delta \Omega_{{\rm c},0} h^2$.
    \label{t:sensitivity_from_data}
  }
  \begin{center}
    \begin{tabular}{c|ccc}
      parameter  & $\Delta p[95\%]$
      & $\left|\frac{\partial\log\Rnu(0.4h/{\rm Mpc})}
      {\partial p}\right|\Delta p$
      & $\left|\frac{\partial\log\Rnu(1.0h/{\rm Mpc})}
      {\partial p}\right|\Delta p$\\
      \hline
      $\Ono h^2$ & $0.0061$  & $0.32$ & $0.69$ \\
      $\sigma_8$ & $0.087$   & $0.10$ & $0.19$ \\
      $n_{\rm s}$ & $0.181$   & $0.019$ & $0.039$ \\
      $h$        & $0.028$   & $0.016$ & $0.037$ \\
      $w_0$      & $0.64$    & $0.013$ & $0.026$ \\
      $w_a$      & $3.0$     & $0.0086$ & $0.042$ \\
      $\Obo h^2$ & $0.00059$ & $0.0041$ & $0.0007$ \\
      $\Omo h^2$ & $0.0053$  & $0.0039$ & $0.017$ \\
    \end{tabular}
  \end{center}
\end{table}

After $\Ono h^2$ and $\Obo h^2$, the next parameters to which $\Rnu$ is most sensitive at $k \gtrsim \kfs$ are $\sigma_8$ and $\Omo h^2$.  At first glance, the relative sensitivities of $\Rnu$ to $\sigma_8$, $\Omo h^2$, and $\Obo h^2$ appear to contradict Fig.~\ref{f:Rnu_first_look}.  However, the range of $\Rnu$ associated with each parameter in that figure is the derivative in Fig.~\ref{f:dlnR_dp} times the parameter range in Table~\ref{t:MT4_range}.  This range for $\sigma_8$ is about six times larger than for $\Omo h^2$, and $100$ times larger than for $\Obo h^2$.

Thus, given a particular data combination, we may define an alternative sensitivity measure for each parameter $p$ by multiplying $\partial \log \Rnu / \partial p$ from Fig.~\ref{f:dlnR_dp} by the range $\Delta p$ allowed by the data.  As an example, we choose the $\nu w$CDM analysis of \citet{Upadhye:2017hdl}, constrained using a combination of CMB, galaxy, and supernova data, and marginalized over a five-parameter model of scale-dependent galaxy bias.  We approximate the $95\%$~CL interval of $\Omo h^2$ by the corresponding one for the CDM alone, its dominant component.  Table~\ref{t:sensitivity_from_data} shows the result at two wave numbers.  While $\Ono h^2$ remains by far the most significant parameter for determining $\Rnu$, $\sigma_8$ is also important.  In summary, while $\Rnu$ is most sensitive to $\Ono h^2$, $\sigma_8$, $\Omo h^2$, and $\Obo h^2$, the first two of these are the most important given current parameter constraints.

%-------------------------------------------------------------------------------
\subsection{Relative clustering in the free-streaming limit}
\label{subsec:enh:relative_clustering_in_free-streaming_limit}

\begin{figure}
  \includegraphics[width=89mm]{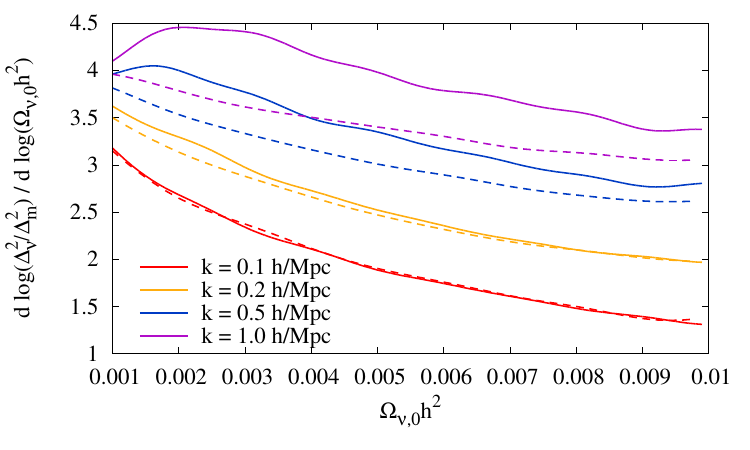}%
  \caption{
    Logarithmic derivative of $\DDnu(k,z)/\DDm(k,z)$ with respect to
    $\Ono h^2$ at several wave numbers $k$, at $z=0$.  Solid lines use
    \cosmicenu{} for $\DDnu$, and dashed lines use \MFLR{}.
    \label{f:dlnRnm_dlnon}
  }
\end{figure}

Next, we consider further the mass scaling of the neutrino power spectrum in the free-streaming limit, raised in the previous subsection.  \citet{Ringwald:2004np} argues that neutrino linear response to non-linear CDM+baryon growth results in a scaling $\DDnu/\DDm \propto \Mnu^4$, while the Tremaine-Gunn bound of \citet{Tremaine:1979we,Shu:1978ApJ...225...83S,Shu:1987ApJ...316..502S,Kull:1996nx} implies  $\partial \log(\DDnu/\DDm) / \partial \log(\Mnu) < 6$.  \citet{Ringwald:2004np} demonstrate using N-body simulations that halos approach this latter bound, and, further, that the bound can be
exceeded, especially in the case of small $\Mnu$,  if one includes all
neutrinos present, rather than only those captured by the halo’s
gravitational potential.

Thus we may expect $\partial \log(\DDnu/\DDm) / \partial \log(\Mnu)$ to rise above four while remaining below six.  Figure~\ref{f:dlnRnm_dlnon} numerically differentiates the ratio of the neutrino power spectrum to the MT4 total-matter power spectrum with respect to $\Ono h^2 \propto \Mnu$ with a step size of $1\%$ in $\Ono h^2 \propto \Mnu$.  Solid and dashed lines respectively use non-linear (\cosmicenu{}) and linear response (\MFLR{}) neutrino power spectra.

Consider first the larger wave numbers.  The \MFLR{} curves approach $4$ from below but never exceed it, as expected.  Meanwhile, the \cosmicenu{} non-linear logarithmic derivatives for $k=0.5~h/$Mpc and $k=1~h/$Mpc both exceed $4$ for small $\Ono h^2$, where these wave numbers are many times the free-streaming scale.  The non-linear enhancement to the mass scaling in \EQ{e:mass-scaling_enhancement} is the difference between the solid and dashed lines.  Focusing on $k=1~h/$Mpc, we find this to be $0.67$ for $\Ono h^2=0.002$ and $0.58$ for $\Ono h^2 = 0.005$.  Since a $3.5\%$ emulator error implies an error of $\sim 0.1$ in the logarithmic derivative, these are consistent with the results of Sec.~\ref{subsec:enh:parameter-sensitivity_of_Rnu}.

This emulator error means that the difference between the linear and non-linear curves for $k=0.1~h/$Mpc is consistent with zero.  The same is true for $k=0.2~h/$Mpc for $\Ono h^2 \gtrsim 0.003$.  Further, the small oscillations observed in some of the logarithmic derivatives are consistent with emulator fluctuations.   Thus Sections~\ref{subsec:enh:parameter-sensitivity_of_Rnu}-\ref{subsec:enh:relative_clustering_in_free-streaming_limit} consistently demonstrate a non-linear enhancement of $\gtrsim 0.5$ to the power law scaling $\partial \log(\DDnu) / \partial \log(\Ono h^2)$ at $k=1~h/$Mpc for $0.002 \lesssim \Ono h^2 \lesssim 0.005$.

%-------------------------------------------------------------------------------
\subsection{Neutrino contribution to the matter power}
\label{subsec:enh:neutrino_contribution_to_matter_power}

The MT4 emulator includes fully linear neutrinos, as implemented in the \cambcode{} code of~\citet{Lewis:1999bs,Lewis:2002ah}, in their CDM+baryon and total matter power spectra, as described in \citet{Saito:2008bp,Agarwal:2010mt,Upadhye:2013ndm,Upadhye:2015lia}.  Since the non-linear clustering of neutrinos increases their power by an order of magnitude relative to linear theory, as shown in Fig.~\ref{f:FFTM_vs_CAMB}, we quantify here the impact of neutrino non-linearity on the matter power spectrum.  Neutrinos will affect $\DDm(k,z)$ in two ways: indirectly, by adding to the gravitational potential, hence enhancing CDM+baryon clustering; and directly, through their inclusion in the total matter power.

Quantifying the indirect effect precisely, by incorporating \FFTM{} into an N-body simulation, is beyond the scope of this study.  However, we may bound this effect.  \citet{Chen:2020bdf} carries out an N-body simulation with neutrino linear response through the \MFLR{} code.  For the largest neutrino fraction considered here, $\Ono h^2 = 0.01$, that study finds an indirect enhancement of $0.05\%$ to the CDM+baryon power spectrum.  Since the non-linear enhancement ratios $\Rnu$ are less than $5$ in Figs.~\ref{f:Rnu_first_look},~\ref{f:Rnu_out-of-sample}, the indirect enhancement is $<0.25\%$.  A more accurate estimate directly multiplying the linear response enhancement of~\citet{Chen:2020bdf} by $\Rnu$ for their model finds an indirect enhancement of $0.16\%$.

\begin{figure}
    \includegraphics[width=89mm]{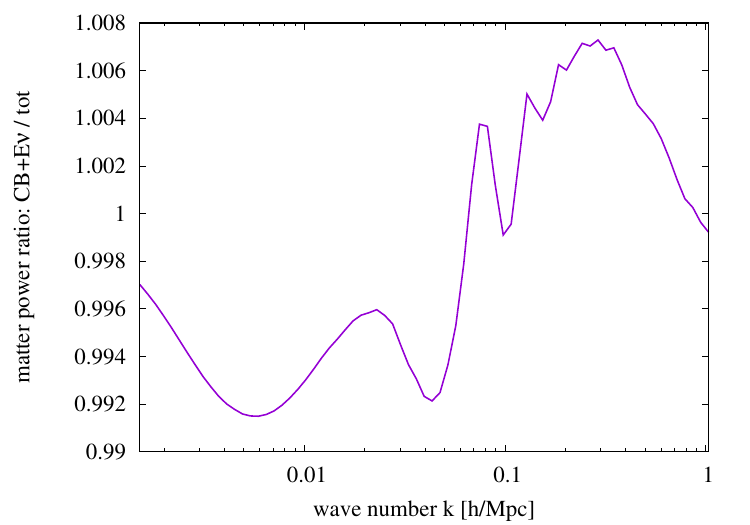}%
    
    \caption{
    Direct non-linear neutrino contribution to the matter power spectrum 
    at $z=0$ for $\Ono h^2 = 0.01$.  $\DDm(k,z)$, computed using the 
    MT4 CDM+baryon power and the \cosmicenu{} neutrino power as in
    \EQ{e:D2m_perfect-corr}, is divided by the MT4 emulated total matter
    power $\DDm(k,z)$.
    \label{f:D2m_enh}
    }
\end{figure}

Figure~\ref{f:D2m_enh} quantifies the direct effect, which is everywhere less than one percent.  Even this is a slight overestimate, as \FFTM{}, hence \cosmicenu{}, assume that CDM, baryon, and neutrino density-contrast monopoles are perfectly correlated.  Under this approximation, the matter power spectrum is 
\begin{equation}
\DDm(k,z) 
= 
\left( \frac{\Ocb}{\Om} \sqrt{\DDcb(k,z)} 
+ \frac{\On}{\Om} \sqrt{\DDnu(k,z)}\right)^2.
\label{e:D2m_perfect-corr}
\end{equation}
\citet{Bird:2018all} shows that the actual neutrino-CDM correlation function drops below unity by $\leq 4\%$ for $k \leq 0.5~h/$Mpc and $\leq 8\%$ for $k\leq 1~h/$Mpc.  The smallness of this deviation is due to the fact that the initially-slowest neutrinos, which contribute the most to small-scale clustering, are also closely correlated with the CDM. This correlation being slightly less than one implies that the actual direct contribution of neutrinos to $\DDm$ is slightly smaller than in Fig.~\ref{f:D2m_enh}.

%%%%%%%%%%%%%%%%%%%%%%%%%%%%%%%%%%%%%%%%%%%%%%%%%%%%%%%%%%%%%%%%%%%%%%%%%%%%%%%%
\section{Conclusions}%%%%%%%%%%%%%%%%%%%%%%%%%%%%%%%%%%%%%%%%%%%%%%%%%%%%%%%%%%%
\label{sec:con}%%%%%%%%%%%%%%%%%%%%%%%%%%%%%%%%%%%%%%%%%%%%%%%%%%%%%%%%%%%%%%%%%

\FFTM{}, the first non-linear perturbative power spectrum calculation for free-streaming particles such as massive neutrinos, provides detailed information on the clustering of neutrinos of different initial momenta.  We have emulated the total non-linear neutrino power spectrum as well as separate power spectra for the ten momentum deciles, each representing a tenth of the neutrino number density.  Our emulated $\DDnu(k,z)$ agrees precisely with \FFTM{} to $<3.5\%$ for $10^{-3}~h/{\rm Mpc} \leq k \leq 1~h/{\rm Mpc}$ and $0 \leq z \leq 3$, as shown in Fig.~\ref{f:D2nu_out-of-sample}.  Individual-decile errors range from about twice as large for the lowest momenta to four times as large for the fastest-moving neutrinos with highly oscillatory density contrasts; see Fig.~\ref{f:D2nuL_outofsample_holdout}.  We have released our emulator as \cosmicenu{}.

Comparing \cosmicenu{} to the highest-resolution simulations of \citet{Euclid:2022qde} in Fig.~\ref{f:codecomp_Mnu}, we found agreement to $3\%$ up to $k=3\kfs = 0.17~h/$Mpc and $19\%$ to $k=0.4~h/$Mpc.  Above this wave number, \cosmicenu{} increasingly underpredicts the simulations of \citet{Euclid:2022qde}, with this underprediction reaching nearly $50\%$ by $k=1~h/$Mpc.  Even this error is not substantially larger than the $30\%-40\%$ scatter between different simulation methods seen in Fig.~\ref{f:codecomp_methods}, so we cannot definitively attribute it either to a non-perturbative effect beyond the capabilities of \FFTM{} or to a systematic error in the simulations.  Importantly, \cosmicenu{} provides a neutrino power spectrum in about ten milliseconds on a standard desktop machine, and we have confirmed that its accuracy is unaffected by rapid variations in the dark energy equation of state.

One strength of the emulation technique is our ability to differentiate numerically the emulated function without the result being dominated by the shot noise and sample variance affecting N-body power spectra.  Section~\ref{sec:enh} took full advantage of this capability by studying the  non-linear enhancement ratio $\Rnu(k,z)$ of \EQ{e:def_Rnu} and the neutrino-to-matter ratio $\DDnu / \DDm$.  Differentiating $\Rnu$ with respect to each of the cosmological parameters, we find that it is most sensitive to the physical neutrino density $\Ono h^2$, but also to $\Obo h^2$, $\sigma_8$, and $\Omo h^2$.  Furthermore, we demonstrated a non-linear enhancement of $\approx 0.5$ to the free-streaming-limit scaling $\partial \log(\DDnu/\DDm) / \partial \log(\Mnu) \rightarrow 4$, meaning that non-linear clustering makes the small-scale density of neutrinos even more sensitive to their mass.  Our results demonstrate the speed and efficacy of the emulation technique in neutrino cosmology.

%%%%%%%%%%%%%%%%%%%%%%%%%%%%%%%%%%%%%%%%%%%%%%%%%%%%%%%%%%%%%%%%%%%%%%%%%%%%%%%%
\subsection*{Acknowledgments}%%%%%%%%%%%%%%%%%%%%%%%%%%%%%%%%%%%%%%%%%%%%%%%%%%%
%%%%%%%%%%%%%%%%%%%%%%%%%%%%%%%%%%%%%%%%%%%%%%%%%%%%%%%%%%%%%%%%%%%%%%%%%%%%%%%%

This project has received funding from the European Research Council (ERC) under the European Union’s Horizon 2020 research and innovation programme (grant agreement No 769130).  Y$^3$W is supported in part by the Australian Research Council’s Future Fellowship (project FT180100031). This research is enabled by the Australian Research Council’s Discovery Project (project DP170102382) funding scheme, and includes computations using the computational cluster Katana supported by Research Technology Services at UNSW Sydney.  The authors are grateful to J.~Conley and S.~Habib for insightful conversations.

%%%%%%%%%%%%%%%%%%%%%%%%%%%%%%%%%%%%%%%%%%%%%%%%%%%%%%%%%%%%%%%%%%%%%%%%%%%%%%%%
\appendix %%%%%%%%%%%%%%%%%%%%%%%%%%%%%%%%%%%%%%%%%%%%%%%%%%%%%%%%%%%%%%%%%%%%%%
%%%%%%%%%%%%%%%%%%%%%%%%%%%%%%%%%%%%%%%%%%%%%%%%%%%%%%%%%%%%%%%%%%%%%%%%%%%%%%%%

%%%%%%%%%%%%%%%%%%%%%%%%%%%%%%%%%%%%%%%%%%%%%%%%%%%%%%%%%%%%%%%%%%%%%%%%%%%%%%%%
\section{Implementation of \cosmicenu{}}%%%%%%%%%%%%%%%%%%%%%%%%%%%%%%%%%%%%%%%%
\label{app:imp}%%%%%%%%%%%%%%%%%%%%%%%%%%%%%%%%%%%%%%%%%%%%%%%%%%%%%%%%%%%%%%%%%

We implement the emulator described in Sections~\ref{subsec:bkg:emulation} and \ref{subsec:emu:emulation_using_sepia} by extracting optimized hyperparameters from {\tt SEPIA}.  Following \citet{Heitmann:2009cu}, we construct a deterministic emulator which uses the mean weights $\bar W_j^{(L)}(\vec C\,)$ of \EQ{e:bar_Wj} as the emulated weights. 

The hyperparameters upon which $\bar W_j^{(L)}(\vec C\,)$ depends may be extracted from {\tt SEPIA}.  For each decile $L$, and for a {\tt SepiaData} object called {\tt data} and a {\tt SepiaModel} object called {\tt model}, the hyperparameter means, basis weights, and basis functions are stored within \SEPIA{} as follows:
\begin{itemize}
\item $w^{*(L)}_{jm}$ in {\tt model.nu.w};
\item $\hat\beta^{(L)}_{j\ell}$ in {\tt model.params.betaU.val};
\item $\lamuh{j}^{(L)}$ in {\tt model.params.lamUz.val};
\item $\lamwh{j}^{(L)}$ in {\tt model.params.lamWs.val};
\item $\mu^{*(L)}_i$ in {\tt data.sim\_data.orig\_y\_mean};
\item $\sigma^{*(L)}$ in {\tt data.sim\_data.orig\_y\_sd}; and
\item $\phi^{(L)}_j(k_i,z_i)$ in the $j$th row, $i$th column of {\tt data.sim\_data.K}.
\end{itemize}

%%%%%%%%%%%%%%%%%%%%%%%%%%%%%%%%%%%%%%%%%%%%%%%%%%%%%%%%%%%%%%%%%%%%%%%%%%%%%%%%
\bibliographystyle{mnras}
\bibliography{CosmicEnu}
\bsp
\label{lastpage}
\end{document}